\documentclass{emulateapj}

\usepackage{amssymb}
\usepackage{graphicx}

\newcommand{\amm}{NH$_{3}$}

\newcommand{\hii}{H\,{\sc ii}}

\newcommand{\cden}{cm$^{-2}$}
\newcommand{\vden}{cm$^{-3}$}

\newcommand{\msun}{M$_{\sun}$}
\newcommand{\kms}{km~s$^{-1}$}

\newcommand{\sfr}{\msun\,yr$^{-1}$}

\shorttitle{ATCA Survey of Ammonia in the Galactic Center}
\shortauthors{Ott et al.}

\begin{document}

%% LaTeX will automatically break titles if they run longer than
%% one line. However, you may use \\ to force a line break if
%% you desire.

\title{ATCA Survey of Ammonia in the Galactic Center: The Temperatures of Dense Gas Clumps between Sgr\,A* and Sgr\,B2}

%% Use \author, \affil, and the \and command to format
%% author and affiliation information.
%% Note that \email has replaced the old \authoremail command
%% frlx om AASTeX v4.0. You can use \email to mark an email address
%% anywhere in the paper, not just in the front matter.
%% As in the title, use \\ to force line breaks.

\author{J\"urgen Ott}
\affil{National Radio Astronomy Observatory, P.O. Box O, 1003 Lopezville Road,
Socorro, NM 87801, USA}
\email{jott@nrao.edu}

\author{Axel Wei{\ss}}
\affil{Max-Planck-Institut f{\"u}r Radioastronomie, Auf dem H{\"u}gel 69,
53121 Bonn, Germany}
\email{aweiss@mpifr-bonn.mpg.de}

\author{Lister Staveley-Smith}
\affil{International Centre for Radio Astronomy Research, The University of Western Australia, 35 Stirling
Highway, Crawley, WA 6009, Australia }
\email{Lister.Staveley-Smith@uwa.edu.au}

\author{Christian Henkel}
\affil{Max-Planck-Institut f{\"u}r Radioastronomie, Auf dem H{\"u}gel
  69, 53121 Bonn, Germany}
\affil{Astronomy Department, Faculty of Science, King Abdulaziz University, P.O. Box 80203, Jeddah, Saudi Arabia}
\email{chenkel@mpifr-bonn.mpg.de}

\and 
\author{David S. Meier}
\affil{New Mexico Institute of Mining and Technology, 801 Leroy Place,
  Socorro, NM 87801, USA}
\affil{National Radio Astronomy Observatory, P.O. Box O, 1003
  Lopezville Road, Socorro, NM 87801, USA}
\email{dmeier@nmt.edu}
%\newpage

\begin{abstract}
We present a large-scale, interferometric survey of ammonia (1,1) and
(2,2) toward the Galactic Center observed with the Australia Telescope
Compact Array (ATCA). The survey covers $\Delta\ell\sim1\degr$
($\sim150$\,pc at an assumed distance of 8.5\,kpc) and
$\Delta b\sim0.2\degr$ ($\sim30$\,pc) which spans the region between
the supermassive black hole Sgr\,A* and the massive star forming
region Sgr\,B2. The resolution is $\sim 20\arcsec$ ($\sim 0.8$\,pc)
and emission at scales $\gtrsim2\,\arcmin$ ($\gtrsim3.2$\,pc) is
filtered out due to missing interferometric short
spacings. Consequently, the data represent the denser, compact clouds
and disregards the large scale, diffuse gas. Many of the clumps align
with the 100\,pc dust ring and mostly anti-correlate with $1.2$\,cm
continuum emission. We present a kinetic temperature map of the dense
gas. The temperature distribution peaks at $\sim 38$\,K with a 
width at half maximum between $18$\,K and $61$\,K (measurements
sensitive within $T_{\rm kin}\sim10-80$\,K). Larger clumps are on average warmer
than smaller clumps which suggests internal heating sources. Our
observations indicate that the circumnuclear disk $\sim1.5$\,pc around
Sgr\,A* is supplied with gas by the 20\,kms\ molecular cloud. This gas
is substantially cooler than gas $\sim3-15$\,pc away from Sgr\,A*. We
find a strong temperature gradient across Sgr\,B2. Ammonia column
densities correlate well with SCUBA 850\,$\mu$m fluxes, but the
relation is shifted from the origin, which may indicate a requirement
for a minimum amount of dust to form and shield ammonia. Around the
Arches and Quintuplet clusters we find shell morphologies with
UV-influenced gas in their centers, followed by ammonia and radio
continuum layers.\\
\end{abstract}

%% Keywords should appear after the \end{abstract} command. The uncommented
%% example has been keyed in ApJ style. See the instructions to authors
%% for the journal to which you are submitting your paper to determine
%% what keyword punctuation is appropriate.


\section{Introduction}
\label{sec:intro}
With a distance of only 8.5\,kpc \citep[IAU value used throughout the
manuscript, but see][who derive $\sim 8.0$\,kpc]{rei09} the Galactic
Center is obviously by far the most nearby core of a massive
Galaxy. It is dominated by a large Galactic Bar potential \citep[$\sim
4.5$\,kpc half-length][]{cab08} and hosts a $\sim
4\times10^{6}$\,\msun\ supermassive black hole
\citep{ghe08,gil09}. The area is at the center of the Galactic
gravitational potential and exhibits some of the most complex and
energetic processes in our Milky Way. The bar forces gas from the disk
to be funneled to the center via $x_1$ and $x_2$ orbits
\citep[e.g.][]{sta04,kim11}. Within the bar potential, most of the
molecular gas resides in a $\sim 500\,$pc region dubbed the Central
Molecular Zone \citep[CMZ; e.g.][]{mor96,oka98,jon12} and can be
considered the region where $x_2$ orbits dominate
\citep[e.g.][]{rod06}. The cloud trajectories in the bar potential are
self-intersecting which results in a large number of cloud-cloud
collisions. The collisions induce shocks that dominate the physical
and chemical properties of the interstellar matter in the region, in
particular the molecular gas. The shocks inject turbulence into the
gas and are a potential heating source. Turbulence is measured by the
increased linewidths of the molecular gas in the Galactic Center
region, with values typically of tens of \kms\ whereas giant molecular
clouds in the disk of the Milky Way exhibit linewidths of only a few
\kms.

Cloud-cloud collisions also influence the properties of star formation
in the region. Studies \citep[e.g.][]{ale07,bar10,lu13} find that the
initial mass function in the central parts of the Galactic Center
region is skewed toward massive stars more than elsewhere in the Local
Universe. In addition, the sheer amount and density of molecular
material (about a few per cent, or a few $10^{7}$\,\msun, of all
molecular gas in the Milky Way is concentrated in about 0.01\% of the
volume) coupled with a star formation rate of $\sim 0.1$\,\sfr\
\citep[e.g.][]{cro11} produces a large number of high mass
stars. Through stellar evolution, they inject mechanical power of
$\sim 10^{40}$\,erg\,s$^{-1}$ in the form of strong stellar winds and
supernovae, and return about 0.02\,\msun\ of material back to the ISM
\citep[for an overview, see][]{cro11}. A dramatic effect of these
energy sources might be the $\sim 10\deg$ sized Fermi bubbles that
extend perpendicular to the Galactic disk far into the Galaxy's halo
\citep[e.g.][]{su10,cro11,car13}.  The high density and short
lifetimes of the massive stars also produce many compact, energetic
objects and the CMZ is under the influence of a large flux of UV and
X-ray photons as well as cosmic rays that can penetrate the molecular
gas. Some of these sources influence gas locally, others influence gas
on larger scales. Overall, however, the state of the gas in the CMZ is
dominated by shock physics \citep[e.g.][]{mar00,mar08,pro08,amo09}.

Given all local and global influences, it is important to understand
how star formation can proceed under these conditions. The molecular
gas defines the site and the material for star formation and it is an
important goal to directly measure the parameters of the molecular gas
such as temperature, density, turbulence, ionization fraction and
magnetic field penetration. In this paper we concentrate on the
temperature of the gas. Our goal is to produce a temperature map of
the Galactic Center gas, in particular of the denser gas phase on
parsec scales that match the size of the individual star formation
locations. A comparison of the temperatures with nearby sources, and a
correlation with the cloud properties will allow us to derive the most
likely heating sources of the gas, and whether they are external or
internal to the gas clouds. 

We decide to use ammonia as a temperature tracer. Ammonia (\amm) has a
peculiar structure of a prolate symmetric top, a tetrahedron with a
nitrogen atom in one corner and three hydrogen atoms in the remaining
corners. The nitrogen atom can tunnel through the plane defined by the
three hydrogen atoms. This possibility removes the degeneracy of the
symmetric top levels ($J$,$K$) at $J>0$, which splits the energy
levels into doublets. Transitions between those levels can be measured
as 'inversion lines'. The inversion line frequencies directly depend
on the rotational states of the molecule. By measuring the intensity
of an inversion line one can thus derive the number of \amm\ molecules
a the given rotational state. The population interchange between
different K ladders is solely governed by collisions, and all
($J>K$,$K$) states decay fast into the metastable $J=K$ levels. The
distribution of all ($J$,$J$) states is thus a result of the
energetics of the collisions, and is simply described by a Boltzmann
distribution at a temperature $T$. Ammonia is thus an excellent
thermometer.

In the following we present interferometric observations of the
ammonia (1,1) and (2,2) transitions across a large fraction in the CMZ
which we use to construct a temperature map of the clumped gas in the
region at pc size resolution. The observations and data reduction are
described in Sect.\,\ref{sec:obs}, followed by our results in
Sect.\,\ref{sec:results}. A discussion of the morphology, kinematics,
temperature structure of the gas clumps and their relation to other
Galactic Center objects are presented in Sect.\,\ref{sec:discuss},
followed by a summary in Sect.\,\ref{sec:summary}.

\section{Observations and Data Reduction}
\label{sec:obs}

\subsection{General Setup}

We observed the Galactic Center during 8 nights on 2005 August 3--8,
10, and 12 with the Australia Telescope Compact Array
(ATCA\footnote{The Australia Telescope Compact Array is part of the
  Australia Telescope which is funded by the Commonwealth of Australia
  for operation as a National Facility managed by CSIRO.}, project id:
C\,1322) with 5 antennas in the most compact, hybrid H\,75 array
configuration (baselines in the range of 31 to 89\,m). Antenna CA\,06 was
also used but, due to the long baselines to the rest of the array
($\sim 4.5$\,km), it is down-weighted and does not contribute
substantially to the surface brightness sensitivity.

The K--band system at the ATCA was tuned to observe both polarizations
at a center frequency of 23.708\,GHz with a bandwidth of 64\,MHz split
into 128 channels. This setup was chosen to observe both, the \amm
(1,1) (rest frequency: 23.6945GHz) and (2,2) (rest frequency:
23.7226\,GHz) lines simultaneously within the same band (when
adjusting for the systemic velocity of the Galactic Center and the
observatory velocity at the time of the observations). This resulted
in a velocity resolution of $\sim 6.3$\,\kms\ at a total Local
Standard of Rest (LSR) velocity range of --180 to +165\,\kms\ common
to both lines. 

Given that the shortest baseline is 31\,m, the array is only sensitive
to emission less than $\sim 1.3\arcmin$, which corresponds to about
half the primary beam of the observations ($\sim 2.4\arcmin$).

\subsection{Mosaicking Technique}
\label{sec:mos}

Our survey covers an area of 0.2\,degree$^{2}$, 1\degr\ in Galactic longitude
($-0.2\degr \leqslant \ell \leqslant 0.8\degr$) and 0.2\degr\ in Galactic
latitude ($-0.13\degr \leqslant b \leqslant 0.07\degr$), i.e., it is
centered on $\ell=0.3\degr$ and $b=0.03\degr$. At the assumed distance of the Galactic
Center, 8.5\,kpc, the covered area
corresponds to widths of $\sim 148$\,pc in $\ell$ and $\sim 30$\,pc in $b$,
$\sim 4440$\,pc$^{2}$ in total.

\begin{figure}
%\epsscale{1} 
%\includegraphics[angle=0,scale=0.5]{gc-scan-dir_paper-small.eps} 
%\vspace{-10cm}
\includegraphics[width=10cm]{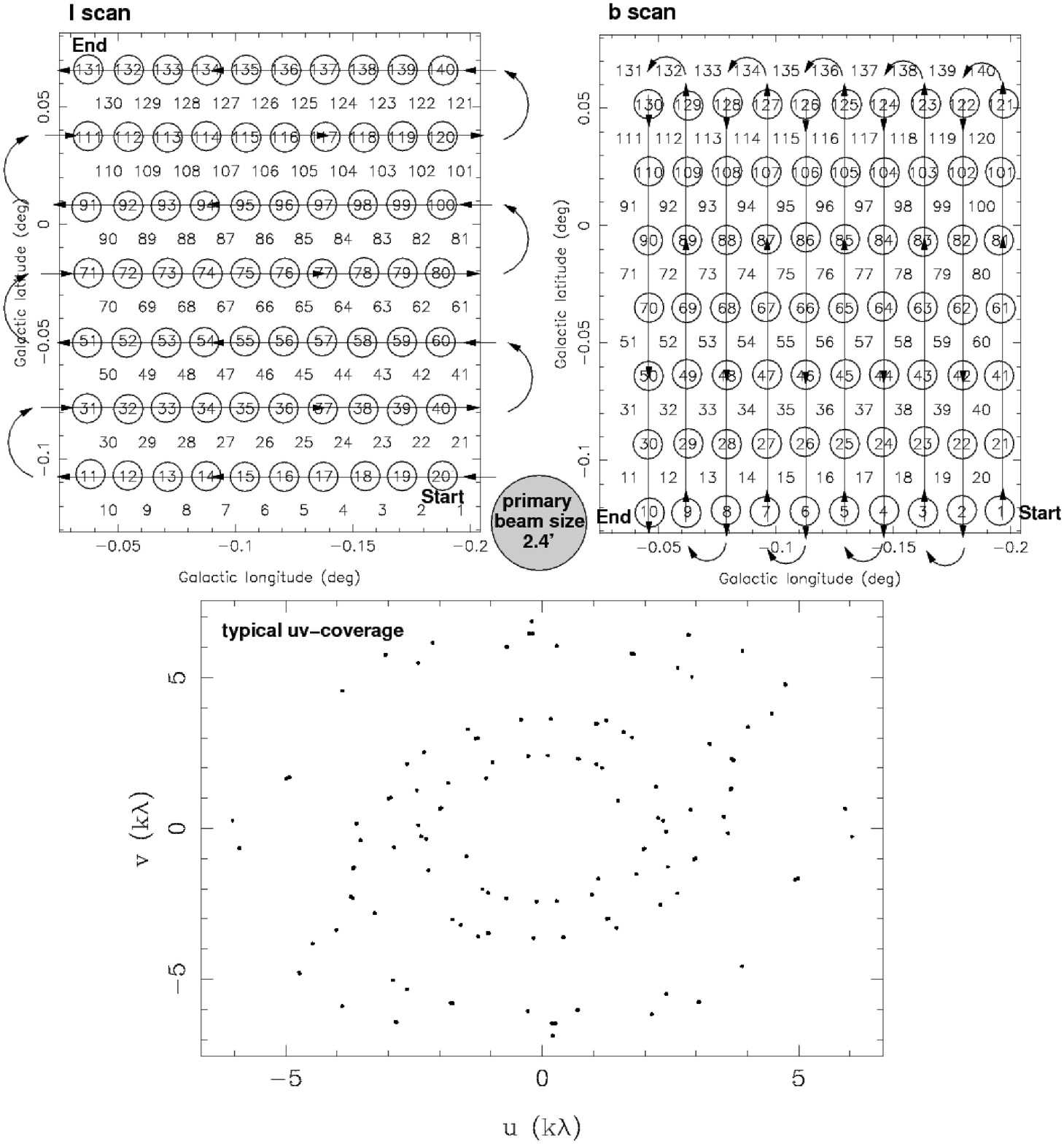} 
%\includeeps[angle=0,scale=0.5]{gc-scan-dir_paper-small.eps} 
%\epsname{gc-scan-dir_paper} 
%%\plotone{f1-small.eps} 
\caption{
{\bf Top:} Scanning pattern of the sub--mosaic fields. Every second row was
scanned first in one direction and later the remaining pointings in the
orthogonal direction. Each number marks a pointing center of the entire
field and numbers in circles indicate those observed in the respective $\ell$
or $b$ scan. The primary beam size of the ATCA in K--band is $\sim
2\farcm4$, shown in between the two panels. {\bf Bottom:} Typical {\sl
uv}--coverage for a single mosaic point. \label{fig:uvcover}}
\end{figure}

A hexagonal, fully--sampled pointing pattern was used to cover the observed
area (primary K--band beam of the ATCA: $\sim 2\farcm4$), 840 pointings in
total. To obtain the most uniform {\sl uv}--coverage we set up the
following mosaicking strategy: The field was split up into 6 different,
equal sized sub--regions along Galactic longitude, with 140 pointing
centers each. In addition, each sub--region was furthermore split up into
two different scans, each of which covers half (i.e. 70 pointings) of the area:
an $\ell$ scan along Galactic longitude and a $b$ scan along Galactic latitude
(see Fig.\,\ref{fig:uvcover}). Scans of the fields were started at
different sidereal times during the 8 nights making sure that each scan and
field, i.e. the entire 840 pointings of the survey, are covered at least
once per night and in rare occasions twice. The typical {\sl uv}--coverage
of a single pointing is shown in the bottom panel of
Fig.\,\ref{fig:uvcover}. Note, however, that due to our distinct $\ell$ and
$b$ scan driving scheme and the overlaps of the primary beams in the
hexagonal pattern, the actual {\sl uv}--coverage at any position on the map
is better. 

Each pointing has an integration time of $\sim 30$\,s per night,
accumulating to $\sim 4$\,min in total. Time lost between the
pointings due to acceleration and deceleration of the ATCA dishes for
our survey amounts to about a second and is therefore negligible.

\placefigure{fig:uvcover}

\subsection{Calibration}
Flux calibration was performed using daily observations of the ATCA
standard calibrator PKS\,1934--638. The $\sim 1.3$\,Jy bright phase
(complex gain) calibrator PKS\,1710--269 was observed twice for 2\,min
each $\ell$ or $b$ scan, i.e. every $\sim 35$\,min (see
Sect.\,\ref{sec:mos}). PKS\,1710--269 is located at the Galactic
coordinates $\ell=357.73044774\degr$, $b=6.99364372\degr$ and can be
considered to be a point source for this array configuration and
frequency. The same source was used to update the pointing of the
individual ATCA antennas every $\sim 1.5$\,hours with typical
corrections of $\sim 5\arcsec$. Antenna--based bandpasses were
determined by daily $\sim 10$\,min integrations on PKS\,1255-053
(also known as 3C279), which has a flux of $\sim 15$\,Jy in K--band. We
estimate the absolute flux uncertainty to $\sim 15$\,\%. Note,
however, that both ammonia lines were observed in the same
band. Relative flux values and resulting parameters such as the
temperature (Sect.\,\ref{sec:temp}) are therefore much more accurate.

All data reduction steps were performed with the {\it MIRIAD} software
package \citep{sau95}. All data were first corrected for atmospheric
opacity (task {\sl atlod} with option {\sl opcorr}) and edited for
corrupt visibilities and shadowed antennas.  Furthermore, the first
seven and last 3 edge channels were discarded. The task {\sl mfcal}
was used to obtain bandpass and complex gain solutions and {\sl
  gpboot} performed absolute flux calibration. This solution was
bootstrapped to the complex gain calibrator (and eventually to the
data) when the latter was at the same elevation as the flux calibrator
PKS\,1934--638. The 1.2\,cm radio continuum was separated from the
{\sl uv} data using line--free channels at either end of the spectra
and in between the \amm\ (1,1) and (2,2) lines (task {\sl uvlin}).

\subsection{Imaging}
The task {\sl invert} was used to Fourier transform the visibilities
into a combined mosaic image of all pointings, including the
corrections for the primary beam shapes. During that step the
visibilities were weighted in the natural scheme as well as by the
inverse of their noise variance. Eventually the data were deconvolved
with the maximum entropy algorithm as implemented in {\sl mosmem}. A
first step comprised a non--restricted 30 iteration run of {\sl
  mosmem} across the mosaic, followed by a restoration of the cube
(task {\sl restor}). This cube, however, still showed considerable
sidelobe structure. The results improved substantially after using the
result from the first application of {\sl mosmem}, convolving it to
about twice the natural beam size, and blanking all values below
100\,mJy\,beam$^{-1}$ per channel, in the following we call this cube
the 'mask'. Eventually, the original data set was deconvolved again,
but we forced {\sl mosmem} to only act on those pixels where the mask
was non--blank. The restoration process of the resulting new
deconvolved model with the residuals of the deconvolution process did
eventually preserve the fluxes in the maps. The effects of the
sidelobes on the data, however, were drastically reduced. We checked
the fluxes of the maximum entropy deconvolution with the more familiar
CLEAN algorithm (task {\sl mossdi}) and found that while the fluxes
are comparable, the extended emission was much better reproduced with
{\sl mosmem}.

The final data have a beam size of $\sim 26\farcs2 \times 16\farcs8$, with a
position angle of $-70$\,\degr, corresponding to a physical resolution in
the Galactic Center of $\sim 1.1$\,pc $\times 0.7$\,pc. The pixel size of
the data was chosen to be 4\arcsec. The flux rms noise per channel is $\sim
7$\,mJy\,beam$^{-1}$ which corresponds to an rms brightness of $\sim
35$\,mK.

Eventually, the total data cube was split into cubes that comprise only the
\amm\ (1,1) or the \amm\ (2,2) inversion lines.

\section{Results}
\label{sec:results}

\subsection{Radio Continuum}
\label{sec:cont}

\begin{figure} 
\epsscale{0.6} 
\includegraphics[angle=90,scale=0.8]{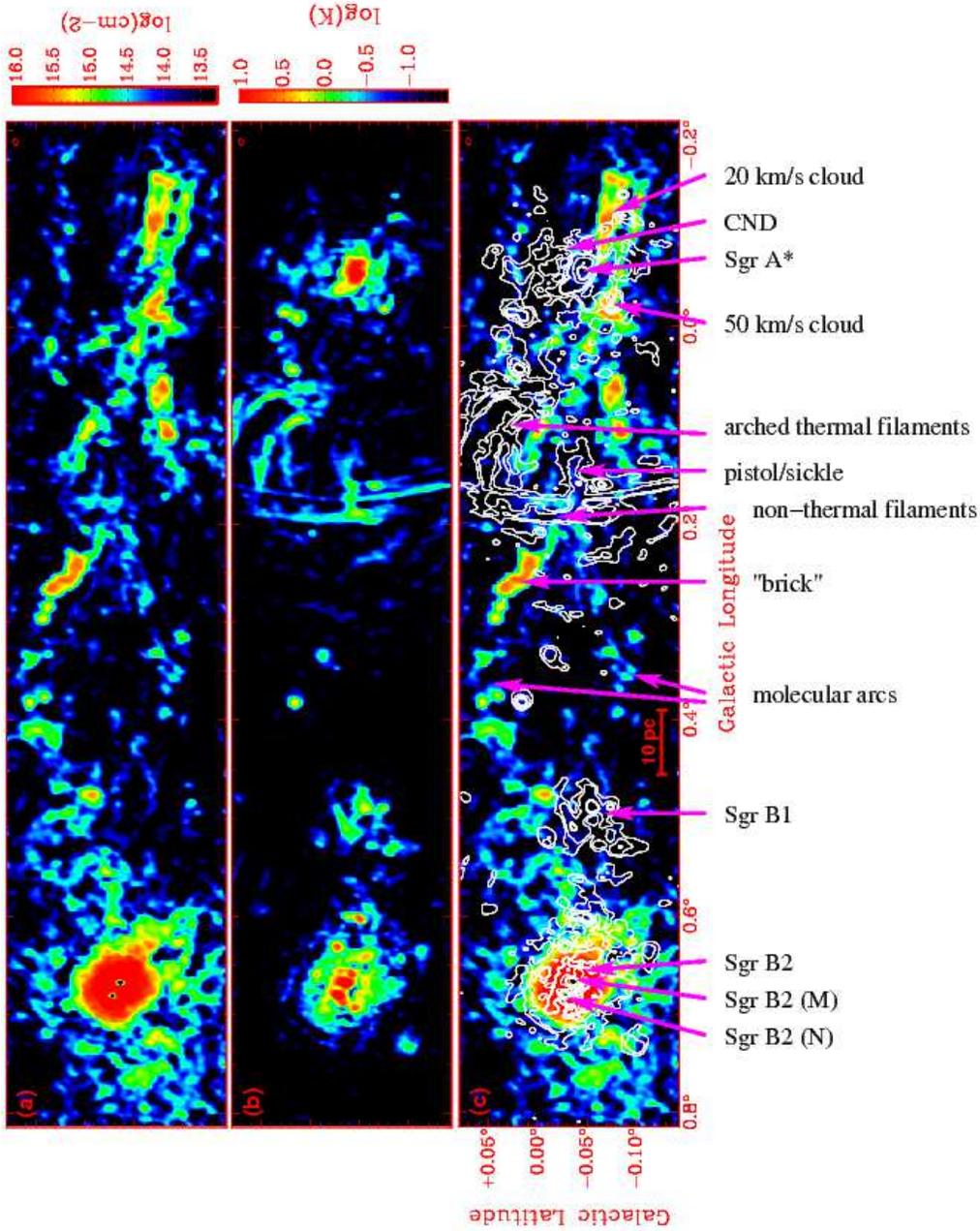} 
%\plotone{cont-pap2-small.eps} 
\caption{{\bf (a)} Integrated intensity map of the \amm\ (1,1)
  emission in logarithmic units of column density. {\bf (b)}
  Logarithmic 1.2\,cm radio continuum. {\bf (c)} The same figure as in
  (a) but with contours of the 1.2\,cm continuum overlaid, the contours
  are in logarithmic units in the range of $-1\leq\log(T_{\rm
    mb;cont})\leq 1$ in steps of 0.5. The beam of the observations,
  ($\sim 26\farcs2 \times 16\farcs8$, PA$=-70\degr$) is shown in the
  upper right corners.  \label{fig:m0}}
\end{figure}

Using the line--free channels from our data, we are also able to
construct a 1.2\,cm continuum map which is displayed in
Fig.\,\ref{fig:m0}(b). The dynamic range across all pointings of the
mosaic is of order $10^{4}$ with the brightest sources toward Sgr\,A*
at $T_{\rm mb}=30.5$\,K (with some confusion from close-by sources),
and Sgr B2 (M) and (N) with main beam brightness temperatures of
48.5\,K and 27.9\,K, respectively. The 'mini--spiral' associated with
Sgr\,A* is also visible as is the Sgr B1 region at
$(\ell,b)\approx(0.49\degr,-0.06\degr)$. Some of the most striking
features are the presence of the non-thermal radio arcs along Galactic
longitudes of $\ell\sim 0.18\degr$. Their fluxes increase from high to
low latitudes from $\sim 0.2$ to $\sim 0.6$\,K with the brightest,
$\sim 1.4$\,K spots in the 'pistol' or 'sickle' region at
$(\ell,b)=(0.189\degr,-0.051\degr)$ \citep[see Fig.\,\ref{fig:m0}; for
a spectral index analysis, see][]{lan99}. The thin, unresolved
vertical filaments appear to be connected in projection to the
thermal, arched filaments that start at $b\approx0.07$ and extend to
$(\ell,b)\approx(0.071\degr,-0.001\degr)$. Between that position and
the Sgr A* region, a number of scattered radio continuum point sources
are detected in the map. Overall, the 1.2\,cm ATCA radio continuum map
exhibits a very similar morphology when compared to the 21\,cm and 90\,cm
Very Large Array maps presented by \citet{yus04}, \citet{ana91}, and
\citet{nor04}. The low frequency maps, however, also contain a number
of finer, arcminute long filaments, perpendicular to the plane,
sometimes called non-thermal radio filaments (NRFs) or threads, that
we do not detect in our 1.2\,cm data.

\placefigure{fig:m0}

\subsection{Sample Spectra}
\label{sec:spec}

\begin{figure} 
\epsscale{0.8} 
\includegraphics[angle=0,scale=0.85]{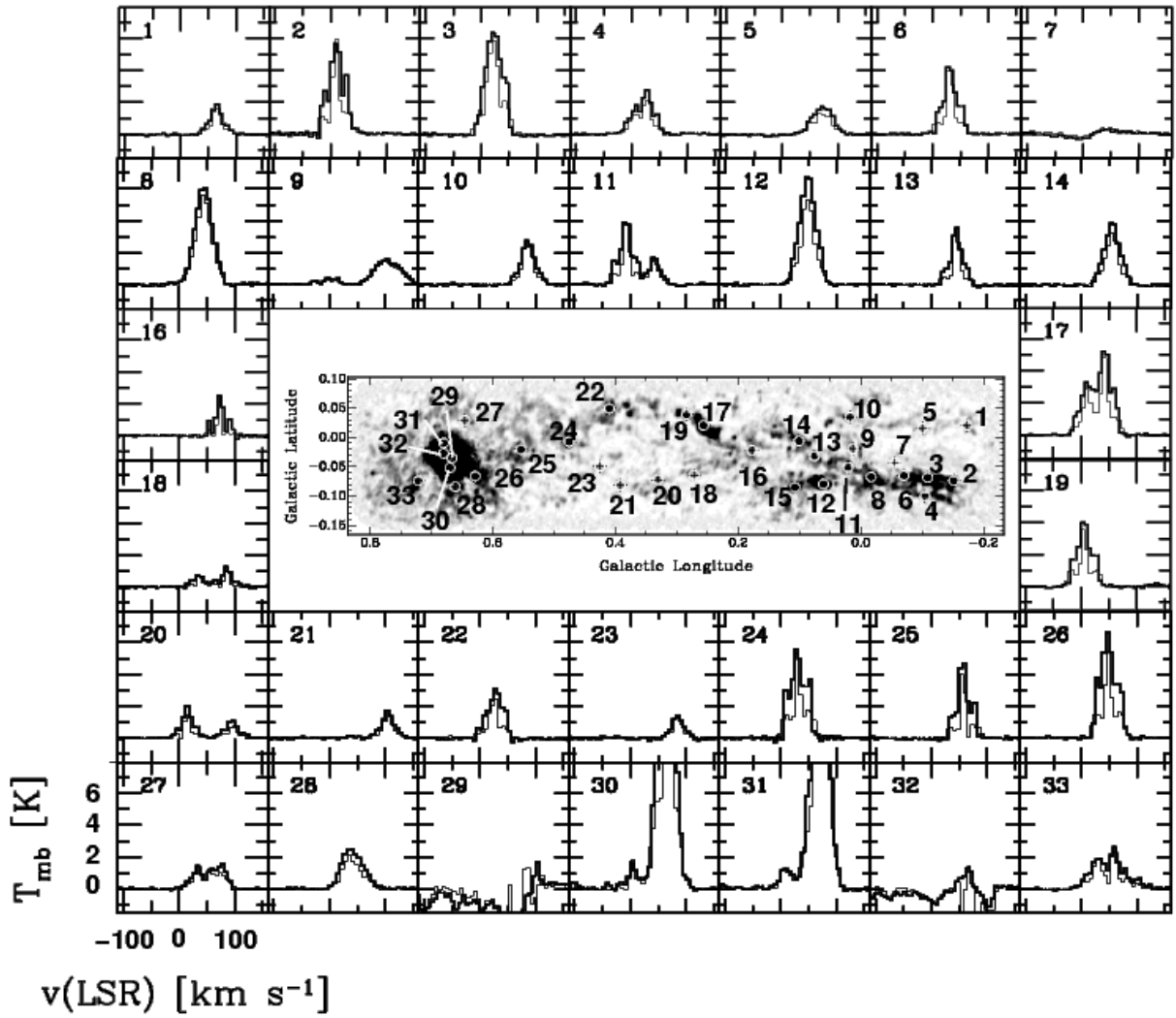} 
%\plotone{moments-pap-small.eps} 
\caption{Ammonia spectra toward individual positions. The numbers in
  the spectral panels correspond to the positions in the map. All
  spectra are at the same scale. \amm\ (1,1) is displayed in {\it
    thick} and \amm\ (2,2) in {\it thin} lines. The parameters of
  the spectra are listed in Tables \ref{tab:pos} and
  \ref{tab:derived}. \label{fig:spec}}
\end{figure}

In Fig\,\ref{fig:spec} we show spectra from the ammonia datacubes at
different positions in the CMZ. The measured parameters of Gaussian
fits to the spectra are listed in Table\,\ref{tab:pos} and derived
values in Table\,\ref{tab:derived}. For better orientation, we also
indicate the area that the positions belong to in
Table\,\ref{tab:derived}. Ammonia (1,1) peak brightness temperatures
are typically around $\sim 1$\,K with values of up to $\sim
17$\,K. \amm(2,2) peak temperatures are typically $\sim 10$\% weaker
than those of the (1,1) line.  FWHM widths of the sample spectra are
20-40\,\kms. Some of the positions show multiple ammonia components
and are listed with ``a'' and ``b'' appendices in the tables. Ammonia has
hyperfine structure which needs to be distinguished from multiple gas
components and also adds to the FWHM. An analysis is presented in
Sect.\,\ref{sec:tau}.

Absorption is observed toward the star forming regions Sgr\,B2 (N) at
position 32, Sgr\,B2 (M) at position 29, and the supermassive black
hole Sgr\,A* at position 7. The absorption spectra are rather complex
and contain all components along the lines of sight from the earth to
the CMZ.

\placefigure{fig:spec}

\subsection{Total Intensity}
\label{sec:m0}

The   integrated   ammonia   (1,1)   intensity   map   is   shown   in
Fig.\,\ref{fig:m0}.  In this  representation, the  velocity integrated
brightness  temperature was  converted \amm\  (1,1)
column densities. The upper level columns of the inversion line can be
derived via the following relation:

\begin{equation}
N_{u}(J,K)=\frac{7.77\times10^{13}}{\nu}\frac{J(J+1)}{K^{2}}\,\,\,\int
T_{\rm mb}\,\, dv
\label{eq:n}
\end{equation}

\noindent \citep[][]{hen00} where $N_{\rm u}$ is given in units of
\cden, the frequency $\nu$ in GHz, the main beam brightness
temperature $T_{\rm mb}$ in K, and the velocity $v$ in \kms. The upper
and lower levels of each inversion line are almost evenly populated
due to the small energy difference. We therefore double each upper
level population to arrive at the total, upper and lower population
for each line.

The dynamic range of the map spans more than 2 orders of magnitude
with the Sgr\,B2 molecular cloud being exceptionally bright. In fact,
the interferometric map around Sgr\,B2 shows a very steep gradient
with a decline in brightness by a factor of $\sim 30$ within a radius
of only 5\,pc \citep[for a more detailed discussion of the profile
see][]{pro08}.  Part of the steep decline, however, might be due to
the missing short spacings which filters out very extended emission
(Sect.\,\ref{sec:obs}). Single dish maps
\citep[e.g.][]{gue81,han06,pur12} show extremely bright and extended
emission, so the short spacing filtering effect is substantial in this
region and we detect only the local peaks of the emission. Toward the
positions of Sgr\,B2 (M) and (N) at $(0.667\degr,-0.036\degr)$ and
$(0.680\degr,-0.029\degr)$, respectively, the ultra--compact \hii\
regions (see also Fig.\,\ref{fig:m0}[b]) are bright enough in the
radio continuum to yield ammonia absorption instead of emission. This
explains the two 'holes' on top of this region.

\begin{figure} 
\epsscale{0.9} 
\includegraphics[angle=90,scale=0.8]{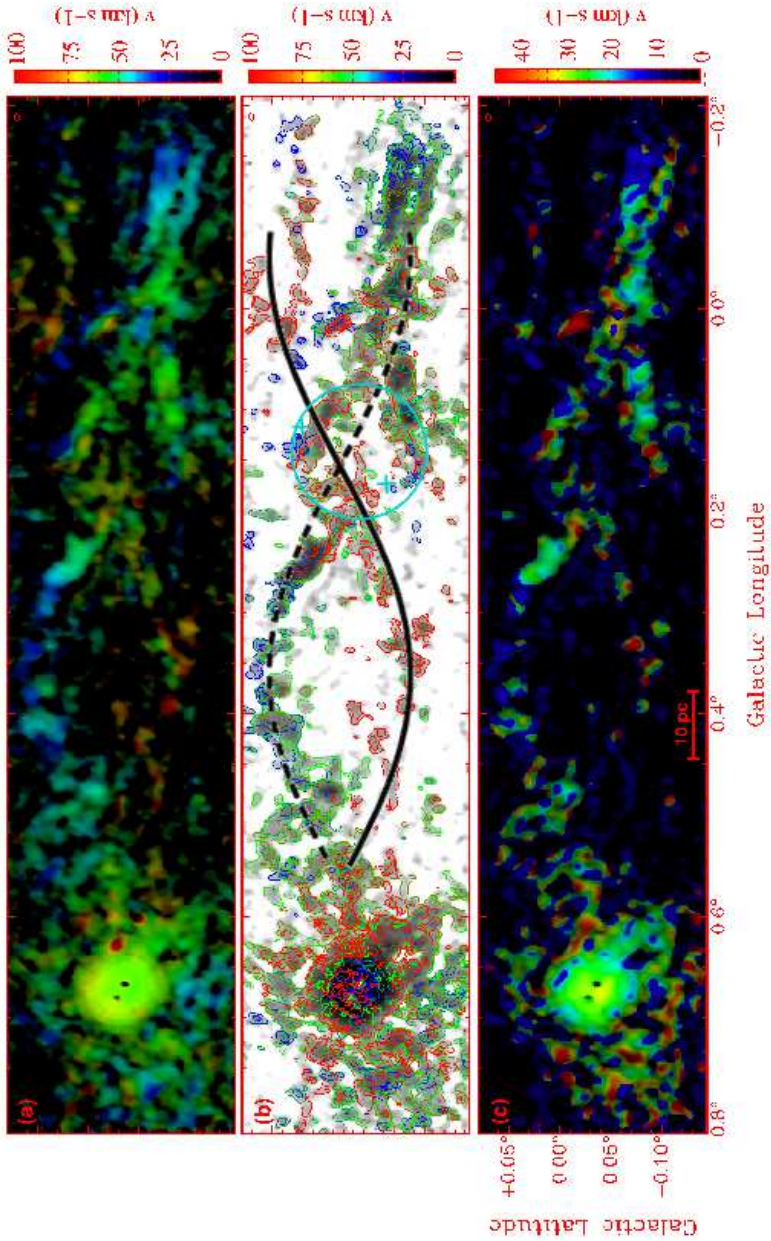} 
%\plotone{cont-pap2-small.eps} 
\caption{The velocity structure as colors projected on the integrated
  emission. {\bf (a)} Intensity--weighted mean velocity map (1$^{\rm
    st}$ moment). {\bf (b)} An intensity contour per plane of the \amm
  (1,1) emission data cube, color coded by velocity (sometimes
  referred to as a 'renzogram'). The plot is overlaid on the
  integrated emission and we indicate two symmetric cosine functions
  that roughly describe the pattern of the molecular arcs. These
  features likely delineate the '100\,pc' ring. The {\it solid} curve
  describes the molecular arc in the foreground and the {\it dashed}
  curve indicates the arc in the background.  The {\it cyan} circle denotes a
  possible cavity close to the site of the Arches and Quintuplet
  clusters, both clusters are marked as {\it cyan crosses}. {\bf (c)}
  velocity dispersion (moment 2). Note that double line spectra may
  show up as artificially high velocity dispersions and also affect
  the mean velocities. \label{fig:vel}}
\end{figure}

Very striking features are the two 'molecular arcs' in between Sgr\,B2
and a Galactic longitude of $\sim 0.2\degr$. The molecular arcs appear
to be symmetric with the symmetry axis slightly below the zero
Galactic latitude line at $b\sim-0.02\degr$; one arc spanning up to $b\sim 0.06\degr$ and
the other up to $b\sim -0.09\degr$ (outlined by a dashed and dotted
line in Fig.\,\ref{fig:vel}). If they are coherent features, they have
lengths of $\sim 0.5\degr$ which corresponds to $\sim 75$\,pc. They
are mostly coincident with the dusty ring detected as a cold structure
in {\it Herschel} temperature maps by \citet{mol11} (see also
Sect.\,\ref{sec:arcs}). One of the brightest clumps being part of the
molecular arcs is the 'brick' G0.253+0.016 at
$\sim(0.253\degr,0.016\degr)$. It is a high density, compact molecular
clump with an integrated brightness temperature of $\sim
230$\,K\,\kms\ that could be the formation site of a massive star
cluster \citep[e.g.][]{lon12}.

\placefigure{fig:vel}

The sightline toward Sgr\,A* is relatively weak with a
temperature-velocity integral of $\sim 11.7$\,K\,\kms. At distances of
$\sim 30\arcsec$ from Sgr\,A* however, we do see surrounding emission
from the circumnuclear disk (CND), a feature that is also prominent in
lines of other molecules such as HCN \citep[][see Sect.\,\ref{sec:cnd}
for details]{chr05,mon09}.

Toward negative Galactic longitudes, the 20\,\kms\ cloud at $(\ell,
b)\sim(-0.1, -0.08)$ is one of the strongest features in our map and
it splits into three almost parallel filaments with lengths of $\sim
0.1\degr$ or $\sim 15\,$pc. The two at higher $b$ are with
110-130\,K\,\kms\ very similar in integrated brightness, whereas the
one at lower $b$ is somewhat weaker with 80-100\,K\,\kms.

In between the 20\,\kms\ cloud and the molecular arcs is the region
where the massive Arches and Quintuplet clusters reside (see
Sect.\,\ref{sec:clusters}). In that part of the CMZ, the molecular
arcs intersect in projection, giving way to a somewhat rounder,
shell-like structure. This structure is located near to a large cloud
that is elongated perpendicular to the Galactic plane in the vicinity
of the non-thermal radio continuum arc (Sect.\,\ref{sec:cont}). From
there, two molecular features are visible that form a ``$>$'' shape
and merge close to the 20\,\kms\ cloud.

\subsection{Global Velocity Structure}

Binning three channels to velocity widths of $\sim 19$\,\kms, the
channel maps of the ammonia (1,1) and (2,2) inversion line emission
are displayed in Figs.\,\ref{fig:chann11} and \ref{fig:chann22},
respectively. As expected, the \amm\ (2,2) emission is generally
somewhat fainter but very similar in structure to the (1,1)
emission. Given the dynamic range of the data (see
Sect.\,\ref{sec:m0}), we display the maps in logarithmic units. At low
velocities, the tip of the 20\,\kms\ cloud as well as the molecular
arc at $b>0\degr$ are visible (see also Fig.\,\ref{fig:m0}). The
southern molecular arc breaks up into clouds at velocities
$\sim$\,20\,\kms\ and $\sim 80$\,\kms\ and is therefore not as
coherent in velocity space as the northern one. It is somewhat
surprising that this arc does not have many cloud components in the
intermediate velocity range at $\sim 50$\,\kms. The molecular gas at
$(\ell<0\degr, b>0\degr)$ is in the velocity range of $50-100$\,\kms\
with an additional component at $\sim -60$ to $-40$\,\kms.  The
intensity--weighted velocity map (moment 1; Fig.\,\ref{fig:vel})
indicates a very similar velocity structure in the $(\ell<0\degr,
b>0\degr)$ region and the arc structure at negative $b$ values between
Sgr\,A* and Sgr\,B2. Note that emission at negative velocities is
relatively weak and thus the intensity--weighted velocity map
displayed in Fig.\,\ref{fig:vel}(b) is biased toward larger values.

\placefigure{fig:chann11}
\placefigure{fig:chann22}

In Fig.\,\ref{fig:pv} we show a position--velocity map of the entire
data cube. The greyscale is the peak intensity determined along
$b$. In this plot the velocity difference between the different
molecular arcs becomes apparent again, with a maximum separation of
$\sim 100$\,\kms\ and an indication that the tips of the arcs arrive
at similar velocities after a smooth velocity difference decrease.

\placefigure{fig:pv}

The mean velocity of Sgr\,B2 is about $\sim 70$\,\kms\ and absorption
of ammonia toward Sgr\,A* is at $\sim 0$\,\kms. Despite local
deviations, a trend can be observed for velocities to decrease from
large $\ell$ toward Sgr\,A*. This trend may be extended to the other
directions. The 20\,\kms\ cloud, for example, is closer to Sgr\,A*
than the 50\,\kms\ cloud which itself is closer to the Center of the
Galaxy than Sgr\,B2 at 70\,\kms.  The velocity decline toward Sgr\,A*
may indicate streaming motions along the central bar of the Galaxy
\citep[e.g., ][]{bli91}, or an x$_2$ orbit along the a 100\,pc dust
ring described by \citet{mol11} (see also Sect.\,\ref{sec:arcs}). In
the first case, if we take the angle of the Galactic bar with respect
to the Sun--Sgr\,A* line of 22\degr\ \citep[][]{bab05}, we derive a
velocity gradient of the molecular gas at the de-projected distance of
Sgr\,B2 ($\sim 270$\,pc) of $\sim 75$\,\kms\ along the bar toward
Sgr\,A* (projected, i.e., measured distance and velocity: $\sim
100$\,pc and $\sim 70$\,\kms). This corresponds to an average velocity
gradient of $\sim 0.28$\,\kms\,pc$^{-1}$. The dynamic timescale for
Sgr\,B2 gas streaming along the bar toward the Galactic Center would
be therefore $\sim 3.5$\,Myr in this picture. If Sgr\,B2 is on the
100\,pc ring (see Sect.\,\ref{sec:arcs}), it will take $\sim2$\,Myr to
reach a position that is aligned in projection with Sgr\,A*.

The linewidths of the ammonia emission are in the range of
20-40\,\kms\, without significant differences toward individual
regions (see Table\,\ref{tab:pos}). In Fig.\,\ref{fig:m0} we show the
2$^{\rm nd}$ moment of the ammonia (1,1) data cube. The very large
values in this map, however, are at positions with multiple line
components and do not reflect the true velocity dispersions of the
ammonia distribution.

\subsection{Temperature Map}
\label{sec:tmap}
One of the main goals of this survey is to derive a temperature map of
the region between Sgr\,A* and Sgr\,B2. As alluded to in the
introduction, the ratio of the upper level populations $N_{u}$ for two
different inversion lines, ($J'$,$J'$) and ($J$,$J$), can be expressed
by a rotational temperature $T_{\rm rot}$.  This parameter can be
derived via the following equation

\begin{equation}
\frac{N_{u}(J',J')}{N_{u}(J,J)}=\frac{g_{\rm op}(J')}{g_{\rm
    op}(J)}\frac{2J'+1}{2J+1}\,\exp\left(\frac{-\Delta E}{T_{\rm rot, JJ'}}\right),
\label{eq:t}
\end{equation}

\noindent \citep[e.g.,][]{ott05}. $\Delta E$ is the energy difference
between the \amm$(J',J')$ and the \amm$(J,J)$ levels in K [$41.2$\,K
between \amm\ (1,1) and (2,2)], $g_{\rm op}$ are the statistical
weights given as $g_{\rm op}=1$ for both para-ammonia (1,1) and (2,2)
inversion transitions. Large Velocity Gradient (LVG) radiative transfer
models \citep[e.g.][]{ott05} show that the population distribution
(and thus the temperature) is almost entirely independent of the
density of the gas, and the above relation can be used for densities
that start at the critical density of ammonia of $\sim 10^3$\,\vden\
up to extreme densities of $\sim 10^7$\,\vden.

For low temperatures, the rotational temperatures correspond well to
kinetic temperatures. At higher energies, however, collisional
de--excitation and the dependency of the collisional cross--section on
the energy result in a deviation from this trend. LVG calculations
suggest that the equation

\begin{equation}
T_{\rm kin}= 6.06\times \exp(0.061\,T_{\rm rot, 12})
\label{eq:tkintrot}
\end{equation} 

provides a reasonable approximation for the conversion from $T_{\rm
  rot, 12}$ to $T_{\rm kin}$ \citep[see figure 5 in][]{ott11}. The relation is
flattening at $T_{\rm kin}\gtrsim 80$\,K ($T_{\rm rot, 12}\sim 40$\,K)
and we therefore consider 80\,K as the lower limit for all higher
temperatures \citep[similar calculations are shown,
e.g. in][]{gue81,wal83,mau86,dan88}. To determine higher temperatures
more accurately, higher $J=K$ inversion lines need to be observed
\citep[e.g.][]{mil13}.

Applying Eq.\,\ref{eq:tkintrot} to our data, we construct a kinetic
temperature map of the region between Sgr\,A* and Sgr\,B2 which is
displayed in Fig.\,\ref{fig:tkin}. Most of the kinetic gas
temperatures, in particular in the molecular arcs between Sgr\,A* and
Sgr\,B2 are relatively uniform at $\sim 30$\,K for smaller clouds and
about $\sim 50$\,K for the larger molecular complexes. Hotter
molecular gas with temperatures of $\gtrsim 80$\,K are found towards
clouds in the vicinity of Sgr\,A* and in the Sgr\,B2 region. In
particular, across Sgr\,B2 a temperature gradient is visible with
temperatures exceeding 80\,K at high Galactic latitude and longitude,
and much colder temperatures of $\sim 30$\,K at lower Galactic
latitudes and longitudes. First indications for this gradient have
been described in the literature e.g. in \citet{chu83} or
\citet{hue93}, but the gradient was never imaged in this lucidity before.

\placefigure{fig:tkin}

\subsection{Optical Depths}
\label{sec:tau}

As already mentioned, the lines in the observed field are, with $\sim
20-40$\,\kms, quite broad. On the other side, our velocity resolution
of $\sim 6.3$\,\kms\ is relatively coarse. Together, both properties
of the data make it difficult to determine the strength of the ammonia
hyperfine lines that are $\sim\pm8$\,\kms\ and $\sim \pm20$\,\kms\
from the main line. However, for some relatively narrow lines it is
possible to detect the hyperfine lines separately and in
Fig.\,\ref{fig:spec} lines at positions 2, 6, 11, 19, 24, 25, and 28,
do show some indications of the $\sim\pm20$\,\kms\ hyperfine
structure. For those lines the main-to-hyperfine line flux ratios are
very consistent at a ratio of $\sim 2.2$ with only $\sim 20$\%
variation. Using:

\begin{equation}
\frac{T_{\rm mb, peak} (\rm main\;lines)}{T_{\rm  mb, peak} (\rm outer\;hyperfine\; lines)}=\frac{1-\exp(-\tau)}{1-\exp(-0.22\,\tau)}
\label{eq:tau}
\end{equation}
\citep{ho83} we derive optical depths $\tau$ to be $\sim 2.4\pm1$.

A spectral fit with the \amm\ fitting routines in the CLASS software
provide similar optical depths, ranging from about $\tau\sim1$ to
$\sim3.5$, with the majority in the $\tau\sim1.5$ to $2$ range. The
fits, taking into account the hyperfine line strengths, also result in
line widths that are smaller than the measured ones and typically
10-20\,\kms\ less than the simple FWHM linewidths.

The calculations of the column densities (Sec.\,\ref{sec:m0}) assume
optically thin ammonia lines. A correction for the sightlines with
moderate optical depths can be expressed by
$\tau/(1-\exp[-\tau])$. For our $\tau\sim1-2.4$ cases this amounts to
$\sim 1.6-2.7$. Changes to the temperature, however, are minimal and
the correction factors cancel out (Eq.\,\ref{eq:t}) when the optical
depths for the (1,1) and (2,2) lines are identical. This is in fact a
good assumption since the average temperatures of the gas is at $\sim
38$\,K (see Sect.\,\ref{sec:temp}, see also Sect.\,\ref{sec:tmap}),
which is in between the energies above ground for the (1,1)
[$E=22.7$\,K] and (2,2) [$E=63.9$\,K] levels, respectively. In the
unlikely edge cases where one of the transitions is optically thin and
the other exhibits $\tau\sim 2$, any temperature corrections would also
be close to a factor of $\sim 2$, with colder temperatures for higher
optical depths of the (1,1) line.

\section{Discussion}
\label{sec:discuss}

\subsection{Comparison to SCUBA observations}
\label{sec:scuba}
In Fig.\,\ref{fig:scuba} we show ammonia (1,1) column density contours
overlaid on a SCUBA 850\,$\mu$m map.  The absolute flux of the SCUBA
data is somewhat uncertain due to the chopping technique used
\citep[see][for a description of the original data]{pie00} and should
be treated as a lower limit since extended flux is likely filtered
out. As mentioned in Sect.\,\ref{sec:mos}, our interferometric ammonia
measurements suffer from similar problems and a short spacing
correction with single dish data would be required to recover the
entire flux of the region. Therefore, on large scales, the comparison
of extended emission in both maps will not be very reliable. On more
compact scales, on which this paper focuses, however, the data can be
compared and the ammonia (1,1) column densities and the SCUBA
850\,$\mu$m data trace each other extremely well.  All features,
except for Sgr\,A* and the ammonia absorption against the compact
\hii\ regions Sgr\,B2(N) and (M) are reproduced in both
datasets. Sgr\,A* is dominated by non--thermal emission \citep[e.g.,
][]{liu01} and is therefore unrelated to dusty molecular gas.

\placefigure{fig:scuba}

To determine the validity of a dust--to--gas relationship between
ammonia and the SCUBA 850\,$\mu$m data, we determined both fluxes in
each pixel (4\,\arcsec\ in size) after smoothing to a common
resolution of 30\arcsec. Following this, we binned the distribution of
the pixels in the \amm\ (1,1) column density versus the SCUBA
850\,$\mu$m fluxes in order to obtain a point density distribution of
these values. The logarithmically scaled point density distribution is
shown in Fig.\,\ref{fig:dustgas}. The tip of the contours (largest
values in dust and gas) follow an almost linear relation as indicated
by the dashed line. The line, however, is not centered on the origin
of the plot but shifted along the SCUBA 850\,$\mu$m axis. This offset
may be the result of the missing large scale flux of both
maps. However, if the 850\,$\mu$m flux is underestimated, the line
would shift even further away from the origin on this axis. On the
other hand, if it is mainly the interferometric ammonia measurements
where we miss flux, the graph would shift up and could be brought in
agreement with starting at the origin. But such a shift would be
extreme as a correction of $\sim 2\times10^{15}$\,cm$^{-2}$ is
required, a value that would almost double the column density for an
average cloud (taking account missing SCUBA flux will increase the
difference even further). Temperature variations of gas and
dust are unlikely to have a significant effect on the slope of
the plot, but could broaden the distribution. We also corrected for
the temperature and calculated {\it total} ammonia column densities from the
(1,1) line and the temperature \citep[via the Eq. A15 in][]{ung86}. As
listed in Table\,\ref{tab:derived}, the (1,1) and (2,2) populations
are together about half of the total ammonia column density that we
derive from our data.  A comparison with the entire ammonia column
density thus leads to a very similar plot as the one shown in
Fig.\,\ref{fig:dustgas} and it does not contain a substantial additive
component. In addition, 850\,$\mu$m dust fluxes are not exceedingly
susceptible to variations in temperature. The offset of the line may
thus be due to a large amount of missing flux from extended emission
mainly of the \amm(1,1) line. A different explanation for the offset
would be that the ammonia forms on dust grains which also shields the
material and that a minimum of dust is required to produce ammonia
efficiently and to protect it against destruction through the
surrounding UV radiation field \citep[see also the discussion
in][]{wei01}.

\placefigure{fig:dustgas}

\subsection{The Radio Filaments and the Molecular Gas}
\label{sec:filaments}
In contrast to the SCUBA maps, the radio continuum emission and the
molecular gas as traced by ammonia show little coincidence. Only the
star forming region Sgr\,B2 exhibits radio continuum emission from the
\hii\ regions at the same positions where large amounts of molecular
gas are found (see Fig.\,\ref{fig:m0}). For this region it is obvious
that the molecular gas most likely feeds the current star forming
activity. As already noted by \citet{meh95}, there is much less
molecular gas in the star forming Sgr\,B1 region at
$(\ell,b)=(0.50\degr,-0.05\degr)$. This may indicate that the Sgr\,B1
region is more evolved than Sgr\,B2 and that the remaining molecular
gas has already been dispersed.

In addition, the non-thermal radio arcs at $\ell\sim0.15$ also occupy
a region with only little ammonia line emission. The pistol/sickle
region at $(0.16\degr,-0.06\degr)$, for example, exhibits \amm(1,1)
column densities of only $\sim 2\times10^{13}$\,\cden. In addition, at
larger Galactic latitudes, the radio arc bifurcates into two roughly
parallel components at $\ell\sim 0.19\degr$ and
$\ell\sim0.17\degr$. Along the filaments, the \amm(1,1) column
densities are $\sim 12\times10^{14}$\,\cden, whereas the column
density in the very thin, long area in between the two filaments rises
up to $\sim 28\times 10^{13}$\,\cden. The radio emission in the two
arched thermal filaments that connect to the main non-thermal arcs and
that stretch from $(0.14\degr,0.05\degr)$ to $(0.08\degr,0.00\degr)$
and from $(0.10\degr,0.08\degr)$ to $(0.06\degr,0.02\degr)$ are also
in regions with low \amm\ average column densities. Toward lower
Galactic latitude and lower Galactic longitudes, however, very
prominent molecular gas clumps are apparent in the integrated
intensity maps (c.f. Fig.\,\ref{fig:m0}). A notable exception to this
behavior is the ``handle'' of the pistol/sickle, the radio filament
that extends roughly perpendicular to the large non-thermal radio arcs
at Galactic latitudes of $\sim -0.04\degr$. Molecular gas appears to
be aligned with this feature and exhibits column densities of $\sim
2.4\times10^{14}$\,\cden. Finally, the region around Sgr\,A* with the
associated mini--spiral is also very bright in 1.2\,cm radio continuum
emission but has only weak features of molecular gas. In particular,
the very prominent circumnuclear, molecular ring around Sgr\,A*, which
is very bright in HCN \citep{chr05}, is not exceptionally strong in
the \amm (1,1) and (2,2) transitions (see Sect.\,\ref{sec:cnd}). The
ammonia feature has a somewhat larger diameter than that of HCN and
column densities of $2-4\times 10^{14}$\,\cden. This is up to an order
of magnitude less than the column densities in the 20 and 50\,\kms\
clouds. The more isolated radio continuum regions at
$(-0.013\degr,0.021\degr)$ and $(-0.053\degr,0.020\degr)$ also show
local peaks in the ammonia distribution of a few
$10^{14}$\,\cden. Associated star formation may thus be in a
relatively early stage.

\subsection{The Molecular Arcs and the 100\,pc Dust Ring}
\label{sec:arcs}

Some of the most prominent features in the interferometric ammonia map
(Fig.\,\ref{fig:m0}) are the two molecular arcs west of Sgr B2,
between $\ell\approx0.2$\degr\ and $0.6$\degr. The two Galactic northern and
southern arcs are almost symmetric in shape, mirrored along a line at
constant $b\approx-0.02$\degr. The northern arc is certainly much brighter
than the low-b counterpart. In Fig.\,\ref{fig:scuba}(c) we show an
overlay of the ammonia to {\it Spitzer} 8\,$\mu$m data. Some of the
detected gas corresponds very well to strong absorption features in
the infrared images. In particular the northern arc is heavily absorbing
the infrared emission whereas the southern arc is hardly visible in
the {\it Spitzer} image at all. Similarly, the string of molecular
clouds north of Sgr A* are less absorbing than those south of Sgr\,A*,
namely the 20 and 50\,\kms\ clouds. This can be interpreted such that
the northern arc and the clouds south of Sgr\,A* are in front of most
of the IR emission whereas the southern arc and the clouds north of
Sgr\,A* are located behind. The different structures are indicated by
the sinusoidal solid (back structure) and dashed (front) lines in
Fig.\,\ref{fig:vel}(b) (also in Fig.\,\ref{fig:pv}).  This picture is
consistent with the morphology of the ``figure 8'' 100\,pc ring that
was recently detected in a {\it Herschel} temperature map by
\citet{mol11}. The shapes of the dust ring and our ammonia arcs are
very similar and the ammonia may thus be emitted by clumps residing in
the dust ring.

\citet{mol11} claim that the ring is rotating, based on single dish
data of the CS molecule. Our data is filtering out the diffuse
emission and the clumps seem to follow a somewhat different
picture. Whereas the \citet{mol11} interpretation suggests that the
foreground arc east of Sgr A* should be redshifted compared to the
background arc, we find the opposite. In Fig.\,\ref{fig:pv} the
northern, foreground arc (dashed line), is at lower velocities
compared the foreground arc (solid line). The velocities of the two
arcs are less distinct toward Sgr B2 and at the position of the
non-thermal arcs at about $b\sim0.15\degr$. This indicates that the
structure might be described by an expansion with a velocity of $\sim
30$\,\kms. Signatures for an expansion were also detected in previous
CO measurements by \citep[][]{sof95}, albeit at much larger expansion
velocities. A comparison of our ammonia data with the
intensity-weighted N$_2$H$^+$ map presented in \citet{jon12}
corroborates the view gained on the basis of the ammonia data alone,
i.e. it appears to contradict the rotating ring picture. Our map,
unfortunately, does not cover the region between Sgr\,A* and Sgr\,C,
which contains the other half of the 100\,pc ring at $\ell<0\degr$.

\subsection{The Temperature Distribution of Molecular Clumps}
\label{sec:temp}

Due to the filtering effect of the ATCA interferometer, we derive the
temperature distribution of molecular clumps more or less directly,
with little contribution from diffuse inter-clump gas.  In
Fig.\,\ref{fig:tkinnh}(a), we show the kinetic temperature
distribution of all pixels in our map. The distribution peaks at about
38\,K and has a FWHM of 43\,K, spanning the 18-61\,K range. We derive
very few areas with a lower limit of kinetic temperatures of
$>80$\,K. Previous studies have detected high temperature components
in the Galactic Center molecular gas using single dish telescopes
\citep[e.g.][]{gue81,hue93b,rod01,ao13,mil13}. It is very likely that
the hot material is mostly diffuse and filtered out by our
interferometric observations in combination with that our (1,1) and
(2,2) lines are not sensitive to high temperatures. This pronounces
the clumpier material which is substantially colder on average.

\placefigure{fig:tkinnh}

From Fig.\,\ref{fig:tkin}(b) we may see a trend that smaller clumps
exhibit colder temperatures. A different view is shown in
Fig.\,\ref{fig:tkinnh} where we plot the logarithm of the total
ammonia column density (cf. Sect.\,\ref{sec:scuba}) to the derived
kinetic temperature of the gas. The pixel by pixel correlation has
been binned and logarithmic contours are shown. The plot confirms that
gas at higher column density also tends to show a higher
temperature. If the heating is external, one would expect it to affect
clouds of all sizes to similar depths. Smaller clumps would thus be more
thoroughly heated whereas larger clumps would be able to maintain a
colder inner core. Since we see the opposite, we may assume that most
of the gas heating takes place internal to the individual clumps. If
the star formation rate per molecular gas mass is roughly constant,
larger clumps will contain more proto-stars, in particular more
massive proto-stars. Those stars are the most likely candidates for
internal heating. We hardly detect massive clumps with very low
temperatures, or small clumps with large temperatures. Small clumps
that do contain massive proto-stars may in fact dissipate relatively
quickly, eluding their observation.

The Galactic Center, however, is known to contain external heating
sources such as cosmic rays, shocks, a strong UV field, X-ray sources,
and dust. As explained above, most of the heating in the clumps is
likely internal. But in some places the external heating may play a
significant role, even for the smaller clumps. Notably, this includes
the bow-shaped region from the thermal, arched filaments, through the
high column density 50\,\kms\ and 20\,\kms\ clouds, in addition to the
compact and dense Sgr B\,2 region which exhibits a strong temperature
gradient (both traced by the yellow and red spots in
Fig.\,\ref{fig:tkin}[a]). The bow-shaped region is close to
Sgr\,A*. Although the CND is cooler (see Sect.\,\ref{sec:cnd}), the
warmer gas may be influenced by activity of Sgr\,A*, which may cause
larger X-ray and cosmic ray fluxes, as well as shocks. PDR effects
may play a major role in the ``mini-starburst'' Sgr\,B2, which is also
a place where the $x_1$ and $x_2$ orbits of the central stellar bar
intersect. The resulting shocks might be an additional heating
mechanism for the molecular gas in that area and can potentially
explain the Sgr\,B2 temperature gradient that is hotter at larger
Galactic longitude values.

\subsection{The Arches and Quintuplet Cluster Regions}
\label{sec:clusters}

Besides the Sgr\,B regions, young stellar clusters are visible all
across the CMZ. Two of the most prominent, massive
clusters are the Arches $(0.121\degr, 0.017\degr)$ and the Quintuplet
$(0.160\degr, -0.059\degr)$ clusters displayed in
Fig.\,\ref{fig:clusters}. The clusters have masses of $\sim
10^{4}$\,\msun\ and are with ages of $\sim 2$\,Myr for the Arches and
$\sim 4$\,Myr for the Quintuple cluster \citep{fig99} relatively
young. The different ages may also be reflected in the dense gas
map. Close to the Arches cluster, a large filament of dense molecular
gas extends roughly parallel to the non--thermal radio continuum
filament. The immediate surroundings of the Quintuplet cluster,
however, are relatively clear of ammonia and may have dispersed over
the lifetime of this cluster. The three--color composite map of the
region in Fig.\,\ref{fig:clusters} shows how different the
distributions of the gas tracers are. CO emission \citep[][]{mar04} is brightest closer
in the direction to the Galactic Center and only little is visible in
a roughly circular region that is defined by the radio arcs, the
pistol/sickle and the thermal arched radio filaments that extend from
the non-thermal arc to smaller Galactic longitudes. This region can be
roughly described by a circle centered on
$(\ell,b)=(0.138\degr,-0.007\degr)$ and having a diameter of
$0.08\degr$ (indicated by the cyan circle in
Fig.\,\ref{fig:vel}[b]). The region, almost entirely surrounded by
radio continuum emission appears to be filled with [\ion{C}{1}]
$^{3}P_{\rm 0}-^{3}P_{\rm 1}$ emitting gas \citep[rest frequency:
492.161\,GHz; data from][]{mar04}. [\ion{C}{1}] is a well--known
tracer for photo--dissociation regions (PDR) and this indicates that
the gas is mainly ionized within this cavity. This is supported by the
detection of Fe\,K$\alpha$ 6.4\,keV emission in the region
\citep{wan02,wan06,tat12}. The gas traced by ammonia is scattered all
across that area. The main filaments, however, are close to the
radio continuum emission toward the inside of the region. The two
stellar clusters are also at the inner edge of the radio continuum
emission. The ammonia channel maps (Fig.\,\ref{fig:chann11}) at this
position show signatures that may be interpreted as an expanding
structure but the relative velocities could well be due to other
motions, too.  If it is expanding, the driving force could be provided
by the gas traced by [\ion{C}{1}]. This scenario would also explain
why the morphology and kinematics of the \amm\ emission does not
follow the overall sinusoidal pattern that is prominently observed by
the molecular arcs between the shell (cyan circle in
Fig.\,\ref{fig:vel}[b]) and Sgr\,B2, which is continued south and
north of Sgr\,A* (Sect.\,\ref{sec:arcs}).

\placefigure{fig:clusters}

The presence of strong [\ion{C}{1}] emission also indicates PDR
conditions and hence a higher ionizing flux and, since the [\ion{C}{1}]
emission is widespread, a higher temperature. The temperature map of
ammonia (Fig.\,\ref{fig:tkin}), however, shows values of $T_{\rm
  kin}\sim 40-60$\,K which are only slightly enhanced compared to
other, more quiet regions of the mapped area. The denser gas therefore
appears to be well--shielded against heating from ionizing flux. If
the picture of an expanding shell structure holds true, the
distribution of the different gas tracers may thus indicate that dense
gas forms a boundary layer that is filled with ionized and PDR gas
(see also the discussion in Sect.\,\ref{sec:filaments}). The dense gas
layer is also where the Arches and Quintuplet clusters are
located. The radio continuum is emitted in an additional sheet
surrounding the dense gas that takes the role of the interface to the
unaffected ISM. Substructures are observed close to the
stellar clusters and within the radio continuum emission. 

The more immediate environment of the Quintuplet cluster is
relatively devoid of ammonia emission but has a local peak in
[\ion{C}{1}]. It also shows radio continuum emission aligned with the
outer part of the structure. This emission forms the pistol/sickle
region. The situation is less clear for the gas in the immediate
surroundings of the Arches cluster. The Arches appears to be located on the
rim of a cavity, that is centered on $(\ell,b)\sim(0.107\degr,0.033\degr)$ and
filled with PDR gas. Since warm, ionized gas follows the path of least
resistance, it may originate from the Arches cluster but then
accumulate in this substructure.

\subsection{The Circumnuclear Disk}
\label{sec:cnd}

The circumnuclear disk around Sgr\,A* is the most central position in
our Milky Way at which substantial amounts of molecular gas are
detected. The ring has been mapped in various dense molecular tracers,
like HCN \citep[e.g.][]{chr05,mon09,mar12} and shows a characteristic diameter
of an arcminute (or about 3\,pc) with a total mass of $\sim
10^{6}$\,\msun\ stored in various clumps \citep{chr05}. A temperature
map with \amm (1,1) column density contours is shown in
Fig.\,\ref{fig:cnd}. We detect a clumpy structure that is in broad
agreement with the HCN observations. \citet{her05} mapped the region
in the ammonia lines with the Very Large Array. They detect an even
clumpier distribution with their more extended array. With our shorter
spacings, we are able to detect the most likely connection of the CND
with the 20\,\kms\ cloud and possibly the 50\,\kms\ cloud \citep[see
also][]{mcg01}. The 20\,\kms\ cloud has a velocity gradient and at the
position where it connects to the CND, the value is about $40$\,\kms\
(see Fig.\,\ref{fig:cnd}). From this level, the values roughly
decrease in a clockwise direction around the CND. This indicates that
gas is fed from the 20\,\kms\ to the CND.

Given that the CND is only $\sim 1.5$\,pc away from the supermassive
black hole Sgr\,A*, it is surprisingly cold. We detect kinetic
temperatures between 25\,K in the clumps and a maximum of $\sim 70$\,K
in some inter-clump regions. Our values are much lower than single
dish measurements that derive a few hundred K for two regions of the
CND \citep[e.g.][]{req12}. This could indicate that the CND is
composed of clumpy cold and diffuse hot material as we are not
sensitive to extended emission.  The temperature structure of the
clumps in the CND is not significantly different from regions further
away, e.g. in the molecular arcs (Sect.\,\ref{sec:arcs}). As seen in
Fig.\,\ref{fig:tkin}, the temperatures of the CND are also lower than
those in a the warmer bow-shaped region described in
Sect.\,\ref{sec:temp}. Cooling and shielding appears to be more
efficient in the CND, if the higher surrounding temperatures are
caused by Sgr\,A*.

\placefigure{fig:cnd}

\section{Summary}
\label{sec:summary}

We observed the region between the supermassive black hole Sgr\,A* and
the massive star forming region Sgr\,B2 in our Milky Way with the ATCA
in the ammonia (1,1) and (2,2) inversion lines ($-0.2\degr \leqslant l
\leqslant 0.8\degr$; $-0.13\degr \leqslant b \leqslant
0.07\degr$). The interferometric data filters the clumpy structure
that otherwise blends with a more diffuse component in single dish
observations. We find the following:

\begin{itemize}

\item The dense gas is distributed in many filamentary structures
  with two prominent molecular arcs between Sgr\,A* and Sgr\,B2. The dense
  molecular arcs are likely located on the '100 pc' ring that was
  recently detected in {\it Herschel} dust temperature maps. 

\item Except for the extreme Sgr\,B2 region, the 1.2\,cm radio
  continuum and ammonia emission are almost anti-coincident. This
  includes the continuum emitted from Sgr\,B1, the non-thermal arcs,
  the arched, thermal filaments, as well as the mini-spiral around
  Sgr\,A*. 

\item SCUBA 850$\mu m$ data trace the dense ammonia gas very well, in
  an almost linear relationship. We see possible indications that a
  minimum amount of dust may be required for substantial amounts of
  ammonia to form and survive the encompassing UV radiation field.

\item A map of kinetic temperatures is constructed that is reasonably
  accurate in the 10-80\,K range. For some of the gas, in particular
  gas close to Sgr\,B2 and in the vicinity of the 20 and 50\,\kms\ clouds
  and the arched, thermal filaments, we derive a lower limit of
  $>80$\,K for the kinetic temperature.

\item The peak of the kinetic temperature distribution is at $\sim
  38$\,K with a FWHM that spans 18-61\,K. On average, larger clumps
  exhibit larger temperatures than smaller clumps. The dominating
  heating mechanism is thus likely internal, and possible sources
  could be proto-stars.

\item The molecular gas for the circumnuclear disk immediately
  surrounding Sgr\,A* with a radius of $\sim 1.5$\,pc is likely being
  supplied by the 20\,\kms\ cloud. The CND is cooler than gas in the
  20\,\kms\ and 50\,\kms\ clouds, and gas closer to the arched,
  thermal filaments.

%\item The molecular gas is with an opacity of typically $\tau\sim2.4$
%  moderately opaque.

\item The region close to the Arches and Quintuplet clusters shows
  signs for a layered structure with more diffuse PDR gas that is
  surrounded by dense ammonia clumps. Furthermore, the radio continuum
  filaments form an outer layer.

\end{itemize}

The Galactic Center is an energetic region and understanding the dense
gas clumps is key to understanding star formation in extreme
conditions. The difference of the diffuse and dense gas components can
be revealed by additional single dish observations. Finally it will be
desirable to observe higher transitions over the full CMZ to obtain a
map of a larger range of temperatures. For example, the \amm(2,2) line
has an energy of $\sim 64$\,K above ground, but the (3,3), (6,6) and
(9,9) levels are at $\sim 123$\,K, $\sim 407$\,K, and $\sim 852$\,K,
respectively, and therefore trace much warmer gas. Other temperature
tracers can be used to calibrate all methods against each other. The
CMZ is also an important benchmark for comparisons with other galactic
nuclei, near and far, active and quiet. Understanding the details of
heating and cooling, star formation, AGN feedback and gas transfer in
the nearby CMZ is therefore indispensable to understand the role of
nuclear regions for galaxy evolution at all cosmic epochs.

\acknowledgements
We thank James Aguirre for providing us access to re-reduced SCUBA data.
The National Radio Astronomy Observatory is a facility of the National Science
 Foundation operated under cooperative agreement by Associated
 Universities, Inc. This research has made use of
NASA's Astrophysics Data System and the NASA/IPAC Extragalactic
Database (NED) which is operated by the Jet Propulsion Laboratory,
California Institute of Technology, under contract with the National
Aeronautics and Space Administration.

\newpage

%\begin{table}
%\rotate
\begin{deluxetable}{r|rrrrrrrrrr}
\tabletypesize{\scriptsize}
%here
%\rotate
\tablecolumns{11}

\tablecaption{Parameters derived from Gaussian fits to the sample
  spectra. Whenever two velocity components could be distinguished,
  they are labeled ``a'' and ``b'' ($\ell$/$b$: Galactic longitude/latitude
  in B1950 coordinates; $T_{\rm mb, peak}$: peak brightness
  temperature; $v_{\rm LSR, peak}$: peak LSR velocity; $\Delta v_{\rm
    FWHM}$: FWHM velocity width, superscripts distinguish between the
  (1,1) and the (2,2) inversion transition lines).\label{tab:pos}}

\tablehead{No. &  \multicolumn{1}{c}{$\ell$}& \multicolumn{1}{c}{$b$}& \multicolumn{1}{c}{$T_{\rm mb,peak}^{11}$}& \multicolumn{1}{c}{$v_{\rm LSR,peak}^{11}$}& \multicolumn{1}{c}{$\Delta v_{\rm FWHM}^{11}$}& \multicolumn{1}{c}{$\int_{11} T_{\rm mb}dv$}& \multicolumn{1}{c}{$T_{\rm mb,peak}^{22}$}& \multicolumn{1}{c}{$v_{\rm LSR,peak}^{22}$}& \multicolumn{1}{c}{$\Delta v_{\rm FWHM}^{22}$}& \multicolumn{1}{c}{$\int_{22} T_{\rm mb}dv$}\\
& \multicolumn{1}{c}{[deg]}& \multicolumn{1}{c}{[deg]}& \multicolumn{1}{c}{[K]}& \multicolumn{1}{c}{[km\,s$^{-1}$]}& \multicolumn{1}{c}{[km\,s$^{-1}$]}& \multicolumn{1}{c}{[K\,km\,s$^{-1}$]}& \multicolumn{1}{c}{[K]}& \multicolumn{1}{c}{[km\,s$^{-1}$]}& \multicolumn{1}{c}{[km\,s$^{-1}$]}& \multicolumn{1}{c}{[K\,km\,s$^{-1}$]}\\}

\startdata

1	&	-0.173	&	0.020	&  $	1.68	\pm	0.05	$ & $	64.31	\pm	0.39	$ & $	27.79	\pm	0.91	$ & $	49.61	\pm	2.16	$ & $	1.56	\pm	0.08	$ & $	64.67	\pm	0.45	$ & $	18.66	\pm	1.06	$ & $	31.02	\pm	2.32	$ \\
2	&	-0.151	&	-0.074	&  $	5.47	\pm	0.55	$ & $	8.37	\pm	1.37	$ & $	28.00	\pm	3.23	$ & $	163.05	\pm	24.91	$ & $	5.80	\pm	0.36	$ & $	9.26	\pm	0.42	$ & $	13.50	\pm	0.97	$ & $	83.41	\pm	7.97	$ \\
3	&	-0.109	&	-0.069	&  $	6.42	\pm	0.13	$ & $	22.90	\pm	0.35	$ & $	37.05	\pm	0.83	$ & $	253.46	\pm	7.54	$ & $	5.19	\pm	0.20	$ & $	21.72	\pm	0.58	$ & $	30.75	\pm	1.37	$ & $	170.05	\pm	10.03	$ \\
4	&	-0.105	&	-0.101	&  $	2.31	\pm	0.08	$ & $	22.04	\pm	0.78	$ & $	43.80	\pm	1.84	$ & $	107.52	\pm	5.96	$ & $	1.62	\pm	0.08	$ & $	20.73	\pm	1.02	$ & $	41.58	\pm	2.41	$ & $	71.82	\pm	5.50	$ \\
5	&	-0.100	&	0.016	&  $	1.72	\pm	0.04	$ & $	72.29	\pm	0.44	$ & $	42.54	\pm	1.04	$ & $	78.11	\pm	2.52	$ & $	1.24	\pm	0.04	$ & $	71.77	\pm	0.70	$ & $	42.46	\pm	1.65	$ & $	56.29	\pm	2.89	$ \\
6	&	-0.070	&	-0.065	&  $	3.75	\pm	0.12	$ & $	31.79	\pm	0.52	$ & $	32.20	\pm	1.23	$ & $	128.56	\pm	6.48	$ & $	4.01	\pm	0.21	$ & $	31.59	\pm	0.37	$ & $	14.57	\pm	0.89	$ & $	62.24	\pm	4.99	$ \\
8	&	-0.017	&	-0.067	&  $	6.06	\pm	0.04	$ & $	44.30	\pm	0.13	$ & $	37.68	\pm	0.29	$ & $	243.15	\pm	2.51	$ & $	5.26	\pm	0.08	$ & $	44.17	\pm	0.24	$ & $	34.17	\pm	0.57	$ & $	191.47	\pm	4.25	$ \\
9a	&	0.013	&	-0.020	&  $	0.37	\pm	0.03	$ & $	0.47	\pm	1.87	$ & $	47.65	\pm	4.45	$ & $	18.92	\pm	2.33	$ & $	0.30	\pm	0.04	$ & $	-0.72	\pm	3.64	$ & $	48.68	\pm	9.01	$ & $	15.30	\pm	3.67	$ \\
9b	&          	&          	&  $	1.56	\pm	0.04	$ & $	104.79	\pm	0.64	$ & $	57.24	\pm	1.53	$ & $	94.92	\pm	3.33	$ & $	1.43	\pm	0.04	$ & $	103.69	\pm	0.70	$ & $	54.99	\pm	1.65	$ & $	83.85	\pm	3.31	$ \\
10	&	0.018	&	0.035	&  $	2.65	\pm	0.04	$ & $	85.96	\pm	0.24	$ & $	30.67	\pm	0.57	$ & $	86.56	\pm	2.14	$ & $	2.39	\pm	0.07	$ & $	85.07	\pm	0.36	$ & $	23.57	\pm	0.85	$ & $	60.08	\pm	2.86	$ \\
11a	&	0.021	&	-0.051	&  $	3.47	\pm	0.14	$ & $	-7.58	\pm	0.62	$ & $	30.67	\pm	1.51	$ & $	113.34	\pm	7.26	$ & $	3.71	\pm	0.22	$ & $	-7.58	\pm	0.62	$ & $	12.54	\pm	0.85	$ & $	49.49	\pm	4.46	$ \\
11b	&          	&          	&  $	1.62	\pm	0.16	$ & $	42.96	\pm	1.20	$ & $	25.04	\pm	2.92	$ & $	43.14	\pm	6.53	$ & $	1.11	\pm	0.13	$ & $	42.96	\pm	1.20	$ & $	36.71	\pm	5.03	$ & $	43.24	\pm	7.81	$ \\
12	&	0.061	&	-0.080	&  $	6.69	\pm	0.06	$ & $	48.66	\pm	0.16	$ & $	35.30	\pm	0.37	$ & $	251.47	\pm	3.49	$ & $	5.09	\pm	0.12	$ & $	48.34	\pm	0.35	$ & $	29.47	\pm	0.83	$ & $	159.68	\pm	5.97	$ \\
13	&	0.075	&	-0.032	&  $	3.23	\pm	0.10	$ & $	45.27	\pm	0.40	$ & $	26.44	\pm	0.93	$ & $	90.87	\pm	4.25	$ & $	3.33	\pm	0.13	$ & $	44.92	\pm	0.26	$ & $	13.52	\pm	0.64	$ & $	47.98	\pm	3.00	$ \\
14	&	0.101	&	-0.006	&  $	3.78	\pm	0.05	$ & $	54.71	\pm	0.22	$ & $	33.77	\pm	0.52	$ & $	136.05	\pm	2.75	$ & $	2.96	\pm	0.09	$ & $	53.96	\pm	0.43	$ & $	29.27	\pm	1.00	$ & $	92.33	\pm	4.18	$ \\
15	&	0.108	&	-0.085	&  $	7.15	\pm	0.11	$ & $	52.06	\pm	0.28	$ & $	35.80	\pm	0.66	$ & $	272.62	\pm	6.61	$ & $	6.13	\pm	0.20	$ & $	51.56	\pm	0.49	$ & $	31.44	\pm	1.16	$ & $	205.18	\pm	10.02	$ \\
16	&	0.178	&	-0.022	&  $	2.45	\pm	0.18	$ & $	73.04	\pm	0.56	$ & $	15.19	\pm	1.32	$ & $	39.64	\pm	4.54	$ & $	2.93	\pm	0.37	$ & $	71.86	\pm	0.42	$ & $	6.61	\pm	0.84	$ & $	20.65	\pm	3.69	$ \\
17a	&	0.257	&	0.020	&  $	2.96	\pm	0.17	$ & $	8.04	\pm	1.26	$ & $	26.86	\pm	2.64	$ & $	84.54	\pm	9.58	$ & $	2.49	\pm	0.19	$ & $	11.84	\pm	0.96	$ & $	24.07	\pm	2.47	$ & $	63.73	\pm	8.18	$ \\
17b	&           &   		&  $	4.86	\pm	0.15	$ & $	43.20	\pm	0.84	$ & $	30.99	\pm	1.86	$ & $	160.22	\pm	10.78	$ & $	4.40	\pm	0.24	$ & $	41.73	\pm	0.43	$ & $	14.63	\pm	1.02	$ & $	68.62	\pm	6.06	$ \\
18a	&	0.272	&	-0.064	&  $	0.62	\pm	0.05	$ & $	36.50	\pm	1.49	$ & $	36.13	\pm	3.71	$ & $	23.68	\pm	3.13	$ & $	0.68	\pm	0.08	$ & $	37.08	\pm	0.62	$ & $	9.77	\pm	1.45	$ & $	7.03	\pm	1.36	$ \\
18b	&           &          	&  $	1.17	\pm	0.07	$ & $	85.72	\pm	0.61	$ & $	20.84	\pm	1.45	$ & $	25.87	\pm	2.33	$ & $	1.03	\pm	0.07	$ & $	85.56	\pm	0.43	$ & $	9.40	\pm	0.71	$ & $	10.33	\pm	1.07	$ \\
19	&	0.284	&	0.039	&  $	3.62	\pm	0.12	$ & $	5.79	\pm	0.56	$ & $	34.57	\pm	1.32	$ & $	133.31	\pm	6.71	$ & $	2.40	\pm	0.17	$ & $	5.66	\pm	1.12	$ & $	31.29	\pm	2.64	$ & $	79.91	\pm	8.90	$ \\
20a	&	0.331	&	-0.072	&  $	1.67	\pm	0.10	$ & $	15.37	\pm	0.78	$ & $	25.90	\pm	1.84	$ & $	46.08	\pm	4.33	$ & $	1.50	\pm	0.10	$ & $	14.91	\pm	0.40	$ & $	11.79	\pm	0.94	$ & $	18.84	\pm	1.99	$ \\
20b	&   		&           &  $	1.00	\pm	0.06	$ & $	93.98	\pm	0.86	$ & $	30.43	\pm	2.04	$ & $	32.32	\pm	2.86	$ & $	0.62	\pm	0.07	$ & $	95.48	\pm	1.71	$ & $	30.50	\pm	4.03	$ & $	20.15	\pm	3.52	$ \\
21	&	0.392	&	-0.082	&  $	1.52	\pm	0.04	$ & $	104.80	\pm	0.40	$ & $	29.40	\pm	0.95	$ & $	47.70	\pm	2.05	$ & $	1.36	\pm	0.05	$ & $	104.63	\pm	0.37	$ & $	20.92	\pm	0.87	$ & $	30.37	\pm	1.68	$ \\
22	&	0.410	&	0.049	&  $	2.95	\pm	0.09	$ & $	28.09	\pm	0.49	$ & $	33.58	\pm	1.15	$ & $	105.41	\pm	4.78	$ & $	2.16	\pm	0.11	$ & $	25.04	\pm	0.62	$ & $	24.10	\pm	1.47	$ & $	55.38	\pm	4.47	$ \\
23	&	0.425	&	-0.050	&  $	1.35	\pm	0.04	$ & $	84.86	\pm	0.38	$ & $	25.91	\pm	0.89	$ & $	37.15	\pm	1.69	$ & $	1.28	\pm	0.07	$ & $	84.69	\pm	0.56	$ & $	20.76	\pm	1.33	$ & $	28.37	\pm	2.39	$ \\
24	&	0.475	&	-0.007	&  $	4.99	\pm	0.23	$ & $	31.73	\pm	0.84	$ & $	37.66	\pm	1.97	$ & $	200.15	\pm	13.84	$ & $	3.05	\pm	0.25	$ & $	32.17	\pm	1.48	$ & $	36.91	\pm	3.49	$ & $	119.81	\pm	14.96	$ \\
25	&	0.554	&	-0.021	&  $	4.01	\pm	0.21	$ & $	57.59	\pm	0.73	$ & $	28.76	\pm	1.73	$ & $	122.97	\pm	9.76	$ & $	4.63	\pm	0.33	$ & $	56.72	\pm	0.27	$ & $	9.71	\pm	0.86	$ & $	47.91	\pm	5.43	$ \\
26	&	0.629	&	-0.066	&  $	5.87	\pm	0.20	$ & $	47.59	\pm	0.61	$ & $	37.16	\pm	1.44	$ & $	232.15	\pm	11.86	$ & $	3.58	\pm	0.23	$ & $	47.94	\pm	1.06	$ & $	33.74	\pm	2.50	$ & $	128.71	\pm	12.62	$ \\
27a	&	0.645	&	0.029	&  $	1.17	\pm	0.07	$ & $	31.94	\pm	1.26	$ & $	28.45	\pm	2.89	$ & $	35.42	\pm	4.20	$ & $	1.04	\pm	0.09	$ & $	31.94	\pm	1.26	$ & $	22.78	\pm	2.26	$ & $	25.14	\pm	3.27	$ \\
27b	&           &           &  $	1.51	\pm	0.07	$ & $	71.31	\pm	1.04	$ & $	31.83	\pm	2.48	$ & $	51.10	\pm	4.59	$ & $	1.04	\pm	0.07	$ & $	71.31	\pm	1.04	$ & $	39.10	\pm	3.27	$ & $	43.21	\pm	4.52	$ \\
28	&	0.661	&	-0.084	&  $	2.48	\pm	0.04	$ & $	40.67	\pm	0.40	$ & $	45.72	\pm	0.93	$ & $	120.65	\pm	3.26	$ & $	1.90	\pm	0.05	$ & $	39.53	\pm	0.56	$ & $	40.40	\pm	1.31	$ & $	81.67	\pm	3.50	$ \\
30a	&	0.670	&	-0.052	&  $	1.48	\pm	0.32	$ & $	4.13	\pm	2.17	$ & $	20.18	\pm	5.13	$ & $	31.71	\pm	10.65	$ & $	1.79	\pm	0.96	$ & $	4.01	\pm	3.28	$ & $	11.19	\pm	6.49	$ & $	21.34	\pm	16.88	$ \\
30b	&          	&   		&  $	16.92	\pm	0.25	$ & $	64.12	\pm	0.25	$ & $	34.55	\pm	0.59	$ & $	622.43	\pm	13.95	$ & $	15.20	\pm	0.67	$ & $	65.64	\pm	0.57	$ & $	26.65	\pm	1.35	$ & $	431.38	\pm	28.88	$ \\
31a	&	0.680	&	-0.010	&  $	1.34	\pm	0.39	$ & $	7.53	\pm	3.16	$ & $	21.73	\pm	7.51	$ & $	30.91	\pm	14.04	$ & $	1.23	\pm	0.39	$ & $	8.35	\pm	3.23	$ & $	20.70	\pm	7.65	$ & $	27.02	\pm	13.18	$ \\
31b	&           &           &  $	11.09	\pm	0.28	$ & $	70.01	\pm	0.53	$ & $	42.52	\pm	1.28	$ & $	502.07	\pm	19.77	$ & $	10.18	\pm	0.28	$ & $	70.50	\pm	0.53	$ & $	38.87	\pm	1.27	$ & $	421.59	\pm	18.10	$ \\
33a	&	0.721	&	-0.074	&  $	0.87	\pm	0.13	$ & $	28.19	\pm	1.07	$ & $	48.93	\pm	7.83	$ & $	45.49	\pm	9.88	$ & $	1.64	\pm	0.14	$ & $	28.18	\pm	1.07	$ & $	25.39	\pm	2.71	$ & $	44.27	\pm	6.02	$ \\
33b	&           &           &  $	1.48	\pm	0.12	$ & $	58.55	\pm	0.45	$ & $	63.72	\pm	5.05	$ & $	100.27	\pm	11.28	$ & $	2.64	\pm	0.21	$ & $	58.55	\pm	0.45	$ & $	11.84	\pm	1.18	$ & $	33.23	\pm	4.21	$ 

\enddata
\end{deluxetable}
%\end{table}

%\begin{table}
%\rotate

\newpage
\begin{deluxetable}{rclrrrrr}
\tabletypesize{\scriptsize}
%\rotate
\tablecolumns{8} \tablecaption{Derived parameters of the sample
  spectra ($N (1,1), N(2,2)$: ammonia column densities of the (1,1)
  and (2,2) lines; $N_{\rm tot}$: total column density for the traced
  ammonia component; $T_{\rm rot,12}$: rotational temperatures based
  on (1,1) and (2,2); $T_{\rm kin}$: kinetic temperatures). Note that
  for rotational temperatures of 40\,K and above, only lower limits to
  the kinetic temperature $>80$\,K can be derived.\label{tab:derived}}
\tablehead{No. & \multicolumn{1}{c}{nearby GC feature} &
  \multicolumn{1}{c}{$N(1,1)$}& \multicolumn{1}{c}{$N(2,2)$}&
  \multicolumn{1}{c}{$N_{\rm tot}$}& \multicolumn{1}{c}{$T_{\rm
      rot,12}$}&
  \multicolumn{1}{c}{$T_{\rm kin}$}\\
  & & \multicolumn{1}{c}{[$10^{14}$\,\,cm$^{-2}$]}&
  \multicolumn{1}{c}{[$10^{14}$\,\,cm$^{-2}$]}&
  \multicolumn{1}{c}{[$10^{14}$\,\,cm$^{-2}$]}&
  \multicolumn{1}{c}{[K]}& \multicolumn{1}{c}{[K]}\\}

\startdata
1	 &dust ring north of Sgr\,A* &  $	6.50	\pm	0.28	$ & $	3.04	\pm	0.22	$ & $	15.4	\pm	1.56	$ & $	32	\pm	2	$ & $	43	^{	+5	}_{	-4	}$\\
2         & 20 km/s cloud & $	21.38	\pm	3.26	$ & $	8.20	\pm	0.78	$ & $	48.78	\pm	16.8	$ & $	28	\pm	2	$ & $	32	^{	+5	}_{	-4	}$\\
3	 & 20 km/s cloud & $	33.24	\pm	0.98	$ & $	16.70	\pm	0.98	$ & $	80.28	\pm	5.66	$ & $	34	\pm	2	$ & $	48	^{	+5	}_{	-5	}$\\
4	 & 20 km/s cloud & $	14.10	\pm	0.78	$ & $	7.06	\pm	0.54	$ & $	34.00	\pm	4.46	$ & $	34	\pm	2	$ & $	47	^{	+6	}_{	-6	}$\\
5	 &dust ring north of Sgr\,A* & $	10.24	\pm	0.34	$ & $	5.54	\pm	0.28	$ & $	25.34	\pm	1.98	$ & $	37	\pm	2	$ & $	55	^{	+6	}_{	-5	}$\\
6	 & 20 km/s cloud & $	16.86	\pm	0.84	$ & $	6.12 \pm	0.250	$ & $	38.24	\pm	4.34	$ & $	27	\pm
2	$ & $	31	^{	+3	}_{	-3	}$\\
7       & Sgr A* &\nodata&\nodata&\nodata&\nodata&\nodata\\
8	 & 50 km/s cloud & $	31.90	\pm	0.32	$ & $	18.82	\pm	0.42	$ & $	81.84	\pm	2.14	$ & $	40	\pm	1	$ & $	66	^{	+5	}_{	-4	}$\\
9a	 & arched thermal filaments& $	2.48	\pm	0.30	$ & $	1.50	\pm	0.36	$ & $	6.44	\pm	1.96	$ & $	41	\pm	5	$ & $	70	^{	+\infty	}_{	-17	}$\\
9b	 & arched thermal filaments& $	12.46	\pm	0.44	$ & $	8.24	\pm	0.32	$ & $	33.88	\pm	3.16	$ & $	45	\pm	2	$ & $	89	^{	+\infty	}_{	-10	}$\\
10	 &arched thermal filaments& $ 11.36	\pm	0.28	$ & $	5.90	\pm	0.28	$ & $	27.72	\pm	1.64	$ & $	35	\pm	2	$ & $	51	^{	+5	}_{	-5	}$\\
11a	 & 50 km/s cloud& $	14.86	\pm	0.96	$ & $	4.86	\pm	0.44	$ & $	33.50	\pm	4.84	$ & $	25	\pm	2	$ & $	28	^{	+3	}_{	-3	}$\\
11b	 & 50 km/s cloud& $	5.66	\pm	0.86	$ & $	4.24	\pm	0.76	$ & $	16.70	\pm	6.96	$ & $	52	\pm	6	$ & $	137	^{	+\infty	}_{	-42	}$\\
12	 & south of the clusters& $	32.98	\pm	0.46	$ & $	15.7	\pm	0.58	$ & $	78.38	\pm	2.56	$ & $	33	\pm	1	$ & $	44	^{	+3	}_{	-3	}$\\
13	 & arched thermal filaments& $	11.92	\pm	0.56	$ & $	4.72	\pm	0.30	$ & $	27.30	\pm	2.90	$ & $	29	\pm	2	$ & $	34	^{	+3	}_{	-3	}$\\
14	 & arched thermal filaments& $	17.84	\pm	0.36	$ & $	9.08	\pm	0.42	$ & $	43.24	\pm	2.08	$ & $	35	\pm	1	$ & $	49	^{	+4	}_{	-4	}$\\
15	 & south of the clusters& $	35.76	\pm	0.86	$ & $	20.16	\pm	0.98	$ & $	90.00	\pm	5.38	$ & $	38	\pm	2	$ & $	60	^{	+6	}_{	-6	}$\\
16	 & non-thermal arcs & $	5.20	\pm	0.60	$ & $	2.02	\pm	0.36	$ & $	11.88	\pm	3.08	$ & $	28	\pm	3	$ & $	33	^{	+6	}_{	-5	}$\\
17a	 & brick & $	11.08	\pm	1.26	$ & $	6.26	\pm	0.80	$ & $	27.92	\pm	7.68	$ & $	38	\pm	3	$ & $	60	^{	+13	}_{	-11	}$\\
17b	 & brick & $	21.02	\pm	1.42	$ & $	6.74	\pm	0.60	$ & $	47.32	\pm	7.18	$ & $	25	\pm	2	$ & $	27	^{	+3	}_{	-3	}$\\
18a	 & southern mol. arc& $	3.10	\pm	0.42	$ & $	0.70	\pm	0.14	$ & $	7.14	\pm	2.24	$ & $	20	\pm	2	$ & $	20	^{	+3	}_{	-2	}$\\
18b	 & southern mol. arc& $	3.40	\pm	0.30	$ & $	1.02	\pm	0.10	$ & $	7.64	\pm	1.56	$ & $	24	\pm	2	$ & $	25	^{	+3	}_{	-3	}$\\
19	 & northern mol. arc& $	17.48	\pm	0.88	$ & $	7.86	\pm	0.88	$ & $	40.98	\pm	4.74	$ & $	31	\pm	2	$ & $	40	^{	+5	}_{	-5	}$\\
20a	 & southern mol. arc& $	6.04	\pm	0.56	$ & $	1.86	\pm	0.20	$ & $	13.60	\pm	2.88	$ & $	24	\pm	2	$ & $	26	^{	+3	}_{	-3	}$\\
20b	 & southern mol. arc& $	4.24	\pm	0.38	$ & $	1.98	\pm	0.34	$ & $	10.02	\pm	2.04	$ & $	32	\pm	3	$ & $	42	^{	+8	}_{	-7	}$\\
21	 & southern mol. arc& $	6.26	\pm	0.26	$ & $	2.98	\pm	0.16	$ & $	14.88	\pm	1.50	$ & $	33	\pm	2	$ & $	44	^{	+5	}_{	-4	}$\\
22	 & northern mol. arc& $	13.82	\pm	0.62	$ & $	5.44	\pm	0.44	$ & $	31.64	\pm	3.26	$ & $	29	\pm	2	$ & $	34	^{	+4	}_{	-3	}$\\
23	 & southern mol. arc & $	4.88	\pm	0.22	$ & $	2.78	\pm	0.24	$ & $	12.34	\pm	1.38	$ & $	39	\pm	2	$ & $	62	^{	+9	}_{	-8	}$\\
24	 & northern mol. arc& $	26.26	\pm	1.82	$ & $	11.78	\pm	1.48	$ & $	61.50	\pm	9.76	$ & $	31	\pm	2	$ & $	40	^{	+6	}_{	-5	}$\\
25	 & tip of mol. arc& $	16.14	\pm	1.28	$ & $	4.70	\pm	0.54	$ & $	36.32	\pm	6.54	$ & $	24	\pm	2	$ & $	25	^{	+3	}_{	-2	}$\\
26	 & Sgr\,B2 area & $	30.46	\pm	1.56	$ & $	12.64	\pm	1.24	$ & $	70.26	\pm	8.18	$ & $	30	\pm	2	$ & $	36	^{	+4	}_{	-4	}$\\
27a	 & Sgr\,B2 area& $	4.64	\pm	0.56	$ & $	2.46	\pm	0.32	$ & $	11.44	\pm	3.22	$ & $	36	\pm	3	$ & $	53	^{	+11	}_{	-9	}$\\
27b	 & Sgr\,B2 area& $	6.70	\pm	0.60	$ & $	4.24	\pm	0.44	$ & $	17.80	\pm	4.08	$ & $	43	\pm	3	$ & $	79	^{	+\infty	}_{	-14	}$\\
28	 & Sgr\,B2 area& $	15.82	\pm	0.42	$ & $	8.02\pm	0.34  $ & $	38.32	\pm	2.48	$ & $	35	\pm	1	$ & $49	^{	+5	}_{	-4	}$\\
29 &Sgr B2 (M)&\nodata&\nodata&\nodata&\nodata&\nodata\\
30a	 & Sgr\,B2 area& $	4.16	\pm	1.40	$ & $	2.10	\pm	1.66	$ & $	10.06	\pm	7.68	$ & $	34	\pm	13	$ & $	48	^{	+58	}_{	-26	}$\\
30b     & Sgr\,B2 area& $	81.64	\pm	1.84	$ & $	42.38	\pm	2.84	$ & $	199.22	\pm	10.70	$ & $	35	\pm	2	$ & $	51	^{	+6	}_{	-5	}$\\
31a	 & Sgr\,B2 area& $	4.06	\pm	1.84	$ & $	2.66	\pm	1.30	$ & $	10.96	\pm	3.00	$ & $	44	\pm	15	$ & $	87	^{	+\infty	}_{	-52	}$\\
31b	 & Sgr\,B2 area& $	64.86	\pm	2.60	$ & $	41.42\pm	1.78 $ & $	174.36	\pm	17.74	$ & $	42	\pm	2	$ & $77	^{	+\infty	}_{	-9	}$\\
32 &Sgr\,B2\,(N)&\nodata&\nodata&\nodata&\nodata&\nodata\\
33a	& Sgr\,B2 area& $	5.96	\pm	1.30	$ & $	4.34	\pm	0.60	$ & $	17.24	\pm	10.04	$ & $	50	\pm	6	$ & $	123	^{	+\infty	}_{	-40	}$\\
33b	 & Sgr\,B2 area& $	13.16	\pm	1.48	$ & $	3.26	\pm	0.42	$ & $	29.86	\pm	7.78	$ & $	22	\pm	2	$ & $	22	^{	+2	}_{	-2	}$
\enddata
\end{deluxetable}

\newpage

\newpage

\newpage

\newpage

\newpage
\begin{figure} 
\epsscale{0.9} 
\plotone{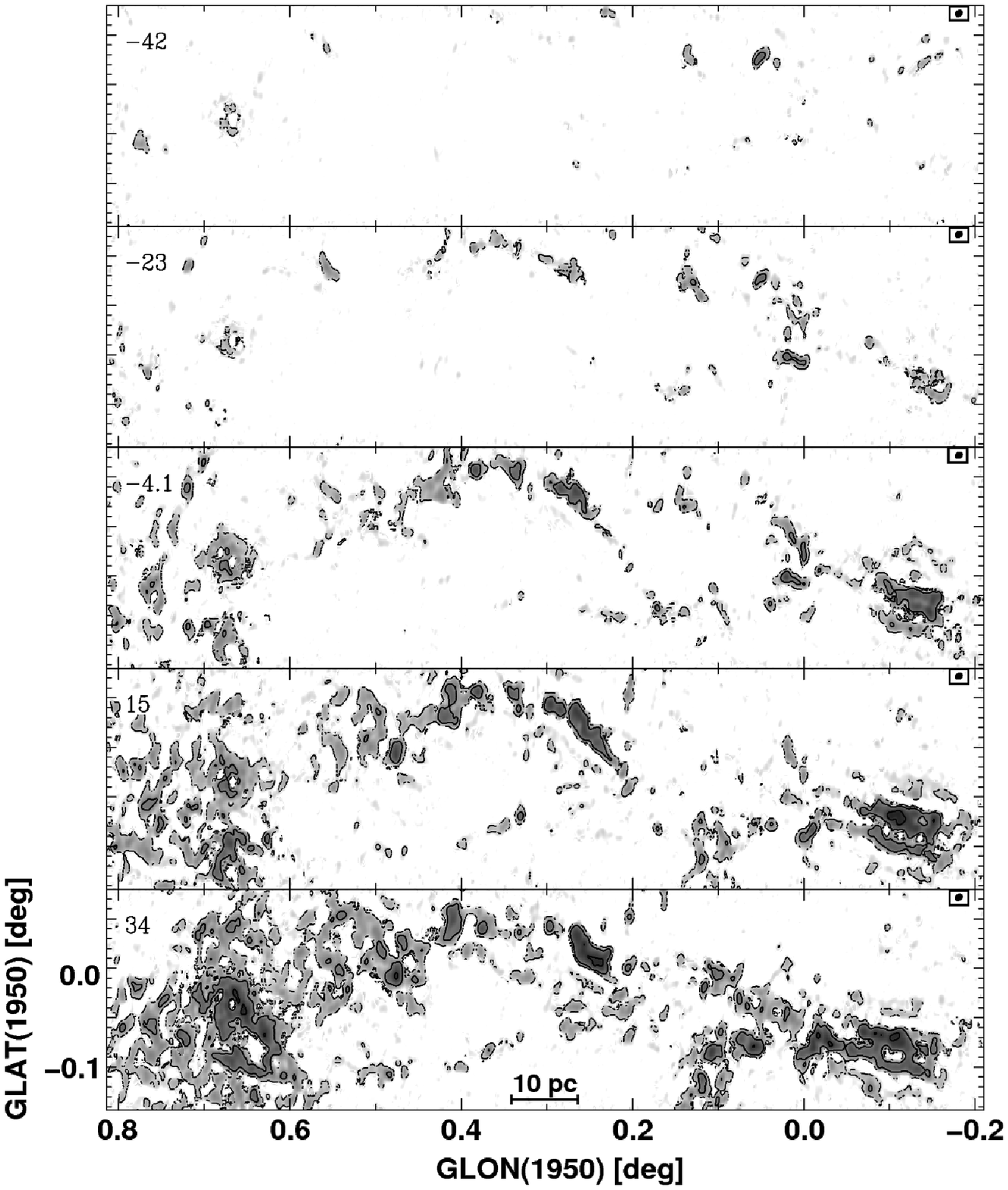} 
%\plotone{f1-small.eps} 
\caption{ Channel maps of the \amm(1,1) emission in logarithmic units
  of $T_{\rm mb}$ (greyscale ranges: $-1.0 \leqslant \log (T_{\rm
    mb}{\rm [K]}) \leqslant 1.0$, contours are at $\log (T_{\rm
    mb}{\rm [K]})$ values of -0.6, 0, and +0.6). Each panel comprises
  three binned channels with mean LSR velocities in \kms\ shown in the
  upper left corners, and the synthesized beam ($\sim 26\farcs2 \times
  16\farcs8$, PA$=-70\degr$) in the upper right
  corners.\label{fig:chann11}}
\end{figure}

\newpage
\addtocounter{figure}{-1}
\begin{figure} 
\epsscale{0.9} 
\plotone{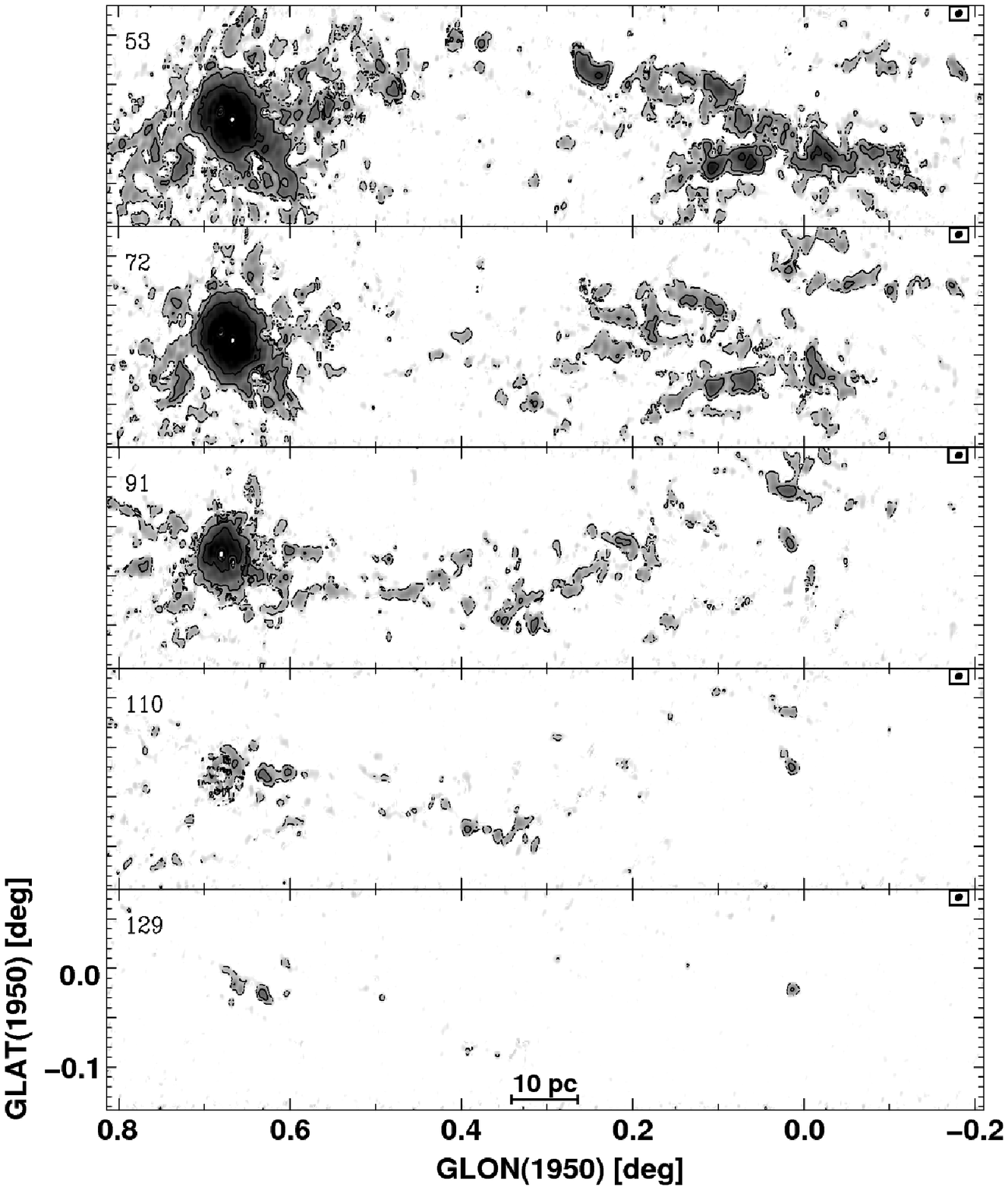} 
%\plotone{f1-small.eps} 
\caption{ continued.}
\end{figure}

\newpage
\begin{figure} 
\epsscale{0.9} 
\plotone{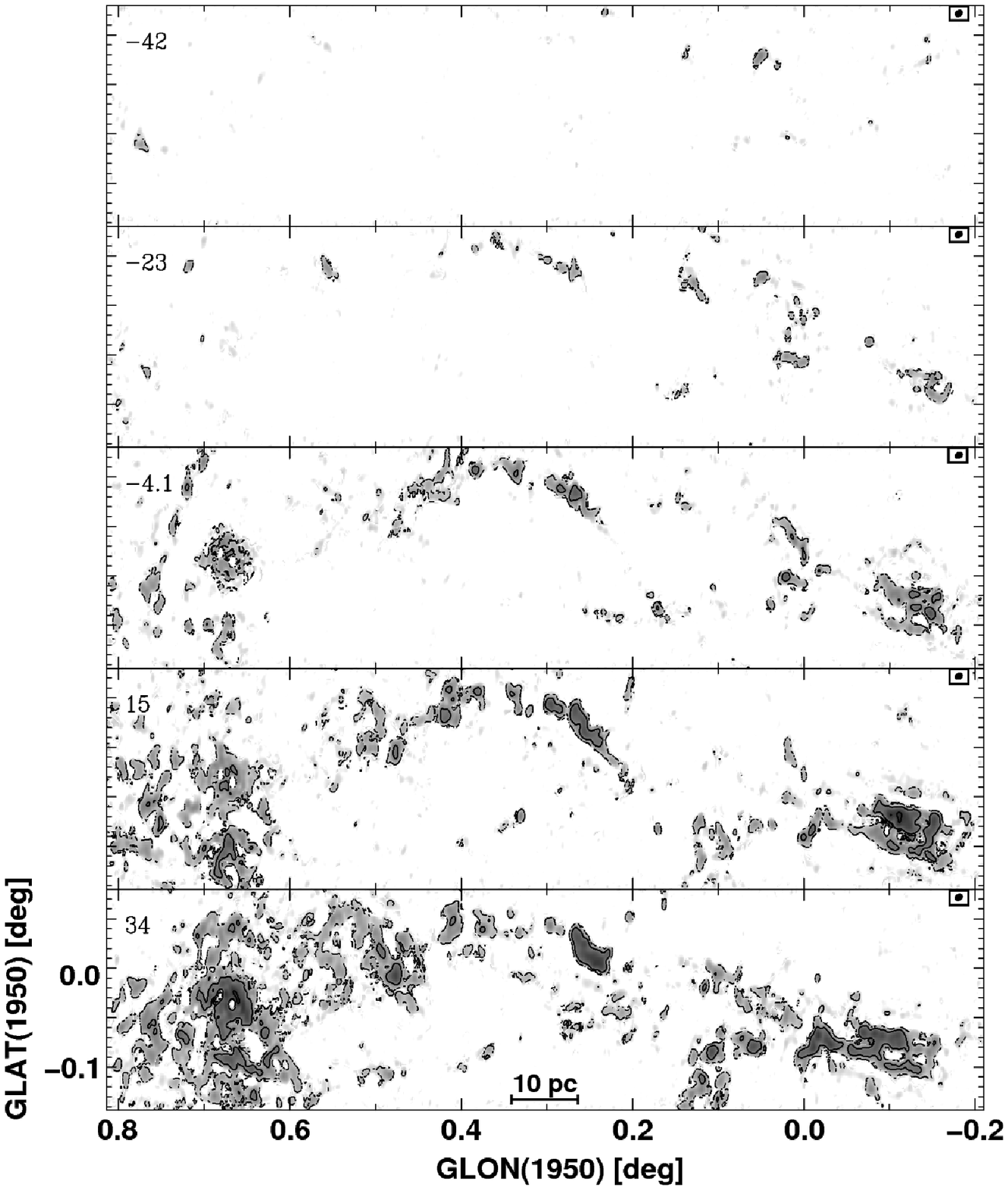} 
%\plotone{f1-small.eps} 
\caption{ Channel maps of the \amm(2,2) emission in logarithmic units
  of $T_{\rm mb}$. Scales, contours, and labels are the same as in
  Fig.\,\ref{fig:chann11}.\label{fig:chann22}}
\end{figure}

\newpage
\addtocounter{figure}{-1}
\begin{figure} 
\epsscale{0.9} 
\plotone{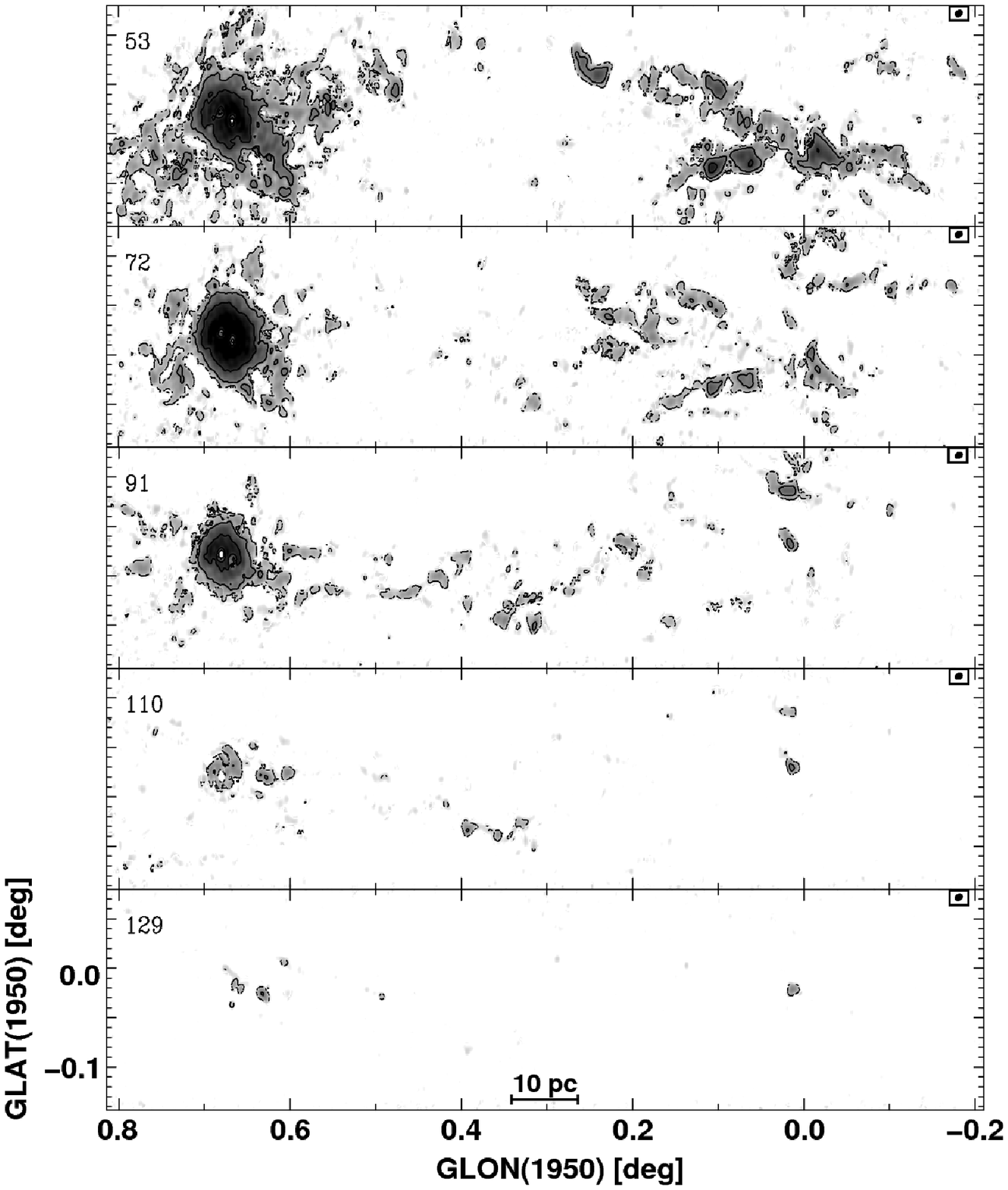} 
%\plotone{f1-small.eps} 
\caption{ continued.}
\end{figure}

\newpage
\begin{figure} 
\epsscale{0.8} 
\includegraphics[angle=0,scale=0.7]{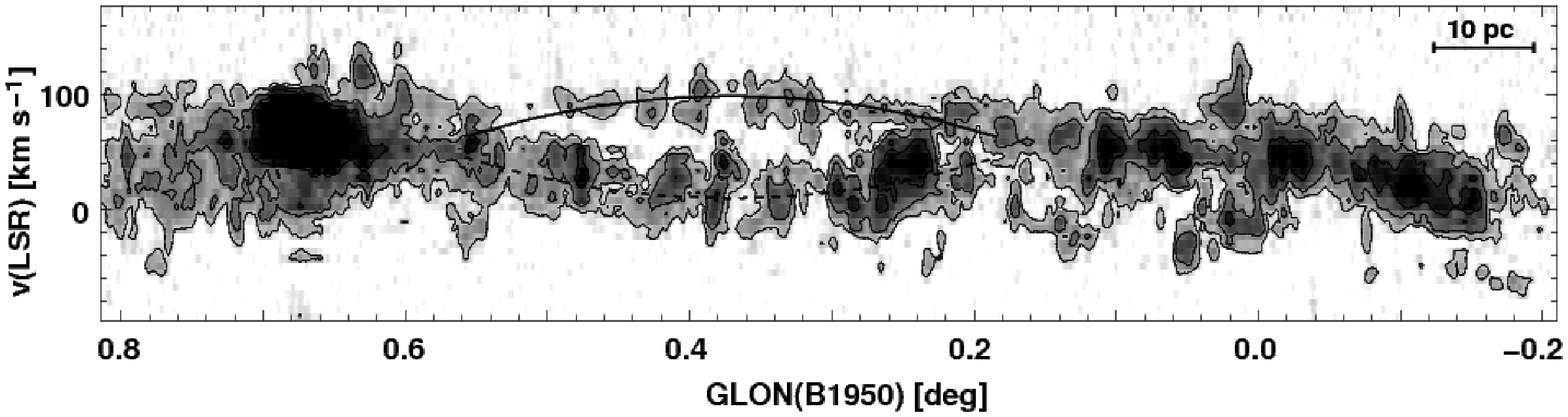} 
%\plotone{moments-pap-small.eps} 
\caption{Position--velocity map of the \amm\ (1,1) emission. The data
  cube for this plot is collapsed across Galactic latitude and the
  peak intensity along that axis is shown in both, greyscale and
  contours. The black solid and dashed lines indicate the same clouds
  as outlined in Fig.\,\ref{fig:vel}(b). \label{fig:pv}}
\end{figure}

\newpage
\begin{figure} 
\epsscale{1.2} 
\includegraphics[angle=90,scale=0.9]{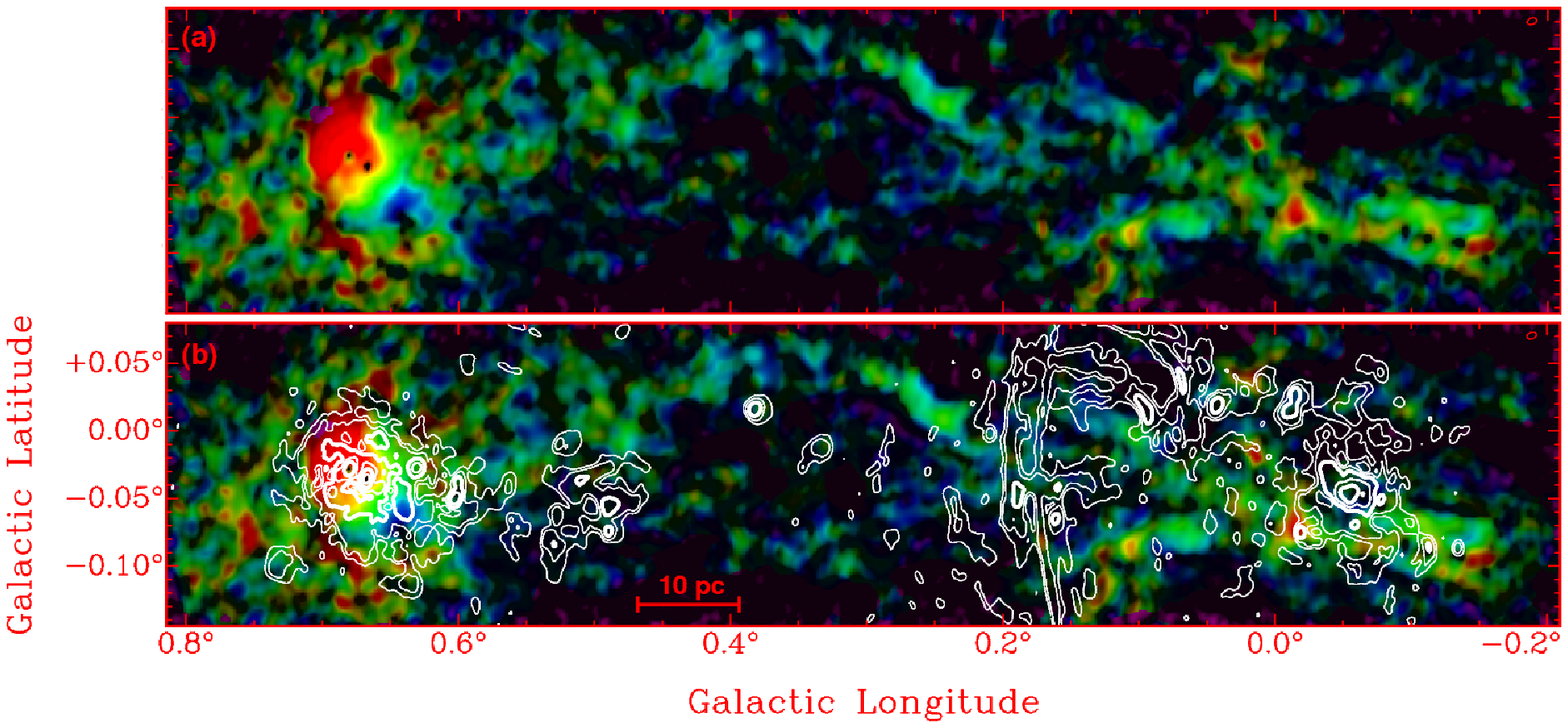} 
 
\caption{Kinetic temperature image of the Galactic Center region
  between Sgr\,A* and Sgr\,B2. Violet: $T_{\rm kin}\lesssim 25$\,K,
  Blue: 25\,K$\lesssim T_{\rm kin}\lesssim 45$\,K, Green:
  45\,K$\lesssim T_{\rm kin}\lesssim 65$\,K Red: $T_{\rm kin}\gtrsim
  65$\,K. In the lower panel, 1.2\,cm continuum contours are overlaid
  with the same spacing as in Fig.\,\ref{fig:m0}. \label{fig:tkin}}
\end{figure}

\newpage
\begin{figure} 
\epsscale{1.2} 
\includegraphics[angle=90,scale=0.9]{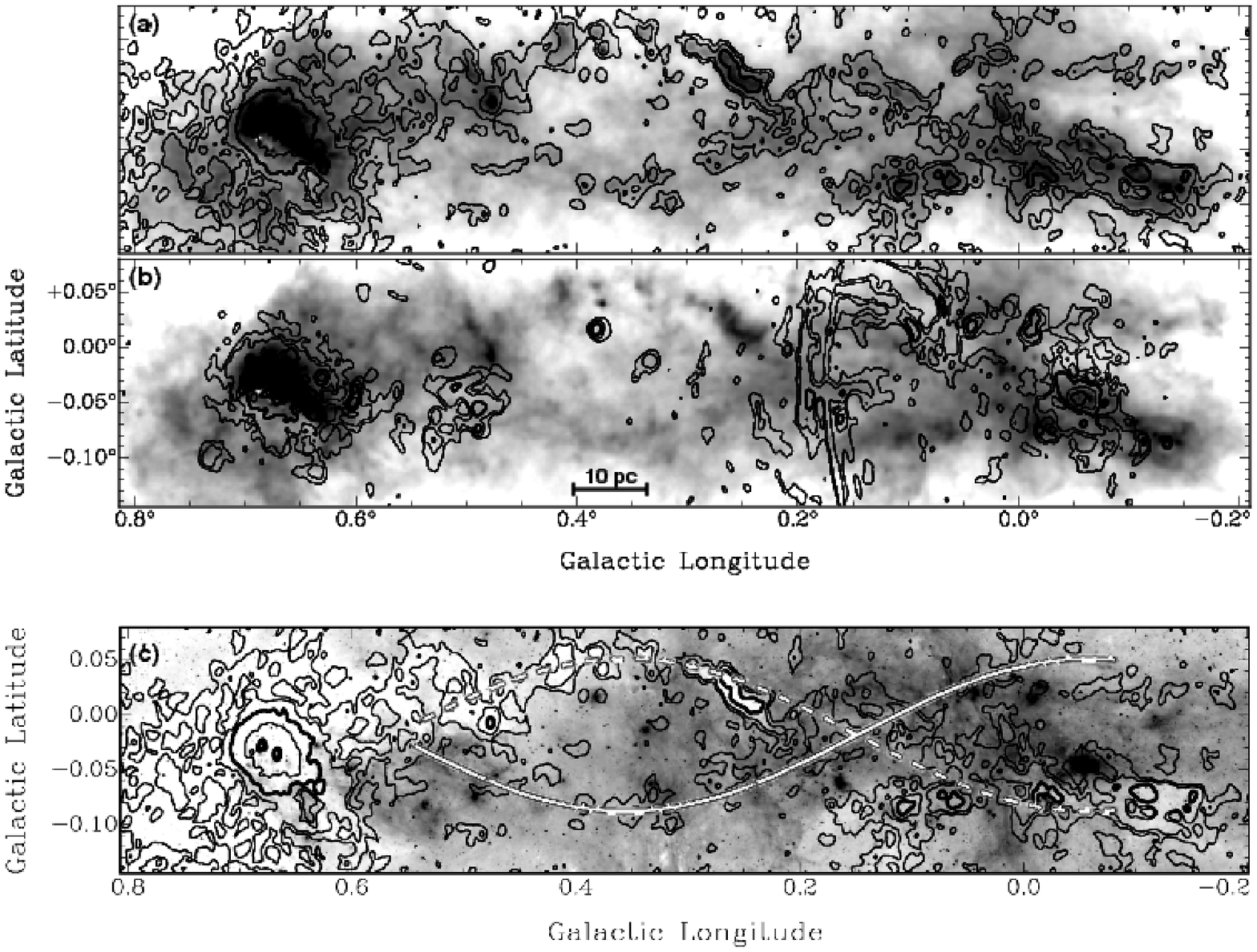} 

\caption{{\bf (a)} SCUBA 850$\mu$m map in logarithmic units with the ammonia
  (1,1) integrated column density overlaid. The contours are within the
  $14\leq\log(N[\rm cm^{-2}])\leq 15.5$ interval, spaced by
  $\log(N[\rm cm^{-2}])=0.5$ {\bf (b)} The SCUBA map with contours of the 1.2\,cm
  continuum spaced as in Fig.\,\ref{fig:m0}. {\bf (c)} {\it Spitzer}
  8\,$\mu$m image in logarithmic units \citep[taken
  from][]{are08}. Overlaid on this image are the ammonia (1,1) column
  density contours and the dotted and dashed white lines correspond to
  those in Fig.\,\ref{fig:vel}[b]. \label{fig:scuba}}
\end{figure}

\newpage
\begin{figure} 
\epsscale{1.2} 
\includegraphics[angle=0,scale=0.8]{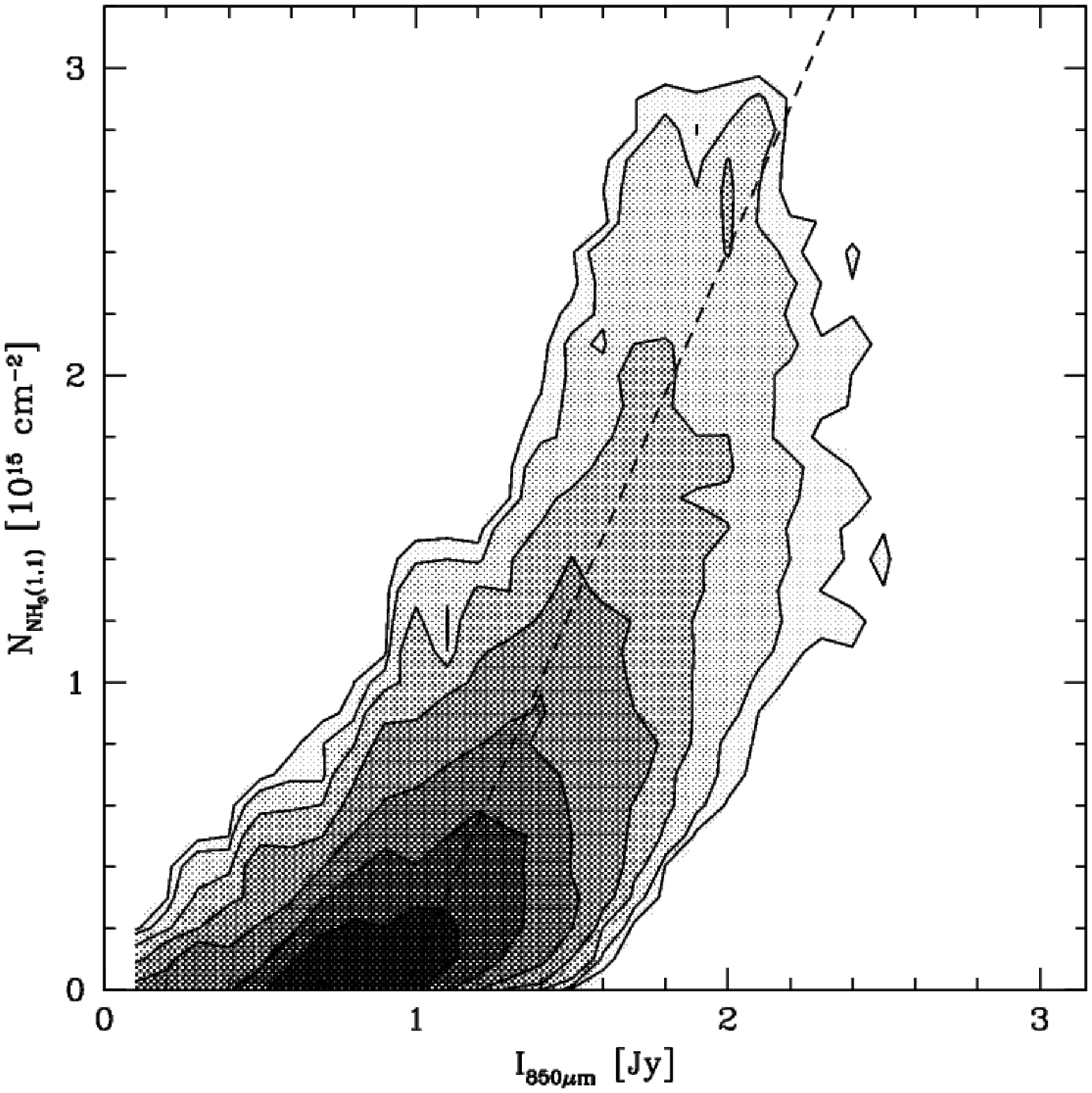} 

\caption{Contour plot of the SCUBA 850$\mu$m fluxes against the \amm\
  (1,1) column densities per pixel at 30\arcsec\ resolution, excluding
  Sgr\,A* and Sgr\,B2. Shown are the contours of the point
  densities in logarithmic units. The tips of the contours
  can be described by an almost linear relationship as is indicated by the
  dashed line. \label{fig:dustgas}}
\end{figure}

\newpage
\begin{figure} 
\epsscale{0.4} 
\includegraphics[angle=0,scale=0.8]{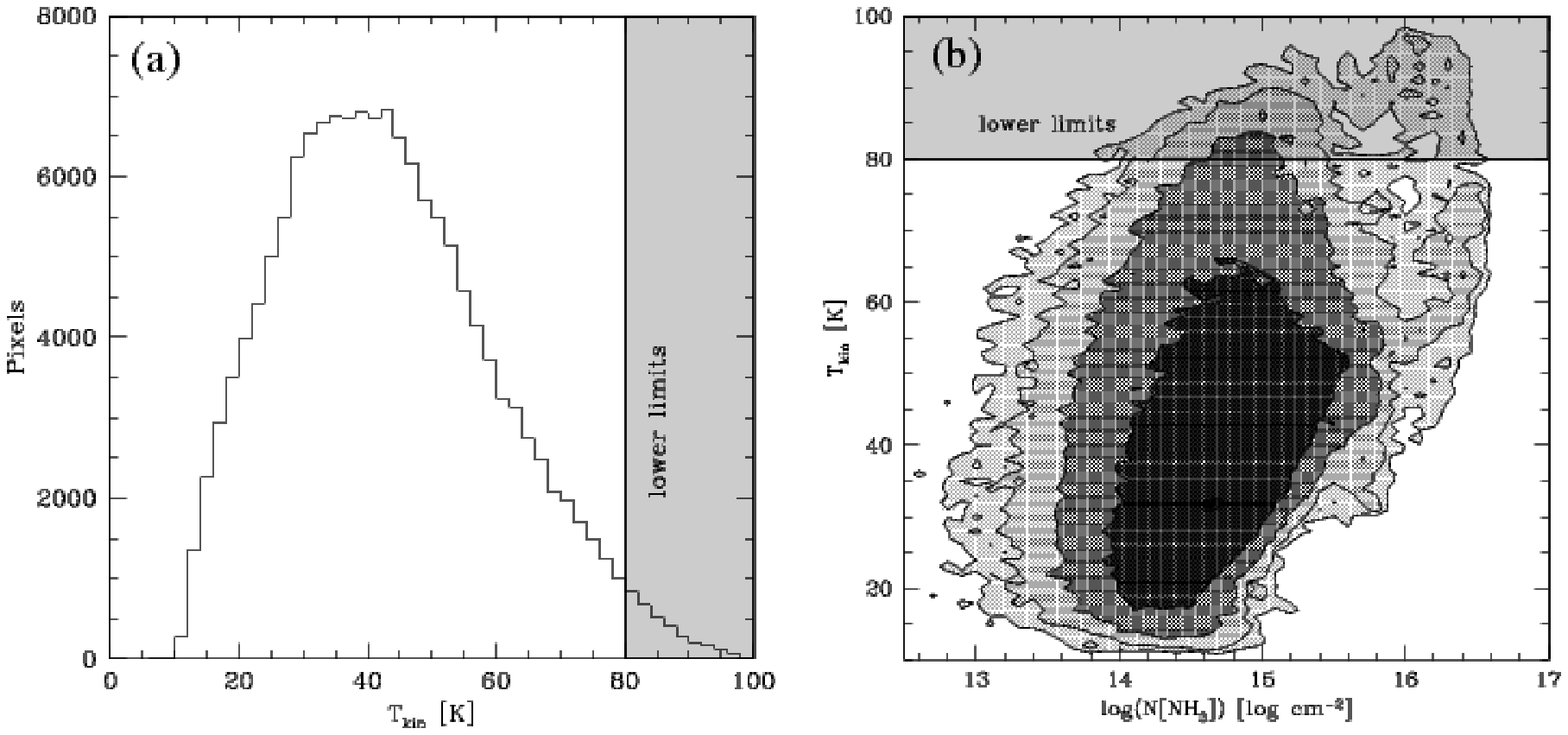} 
\caption{(a) The distribution of kinetic temperatures in our map. (b)
  Kinetic temperatures as a function of \amm\ column density per
  pixel. All temperatures above 80\,K need to be considered as lower
  limits. \label{fig:tkinnh}}
\end{figure}

\newpage
\begin{figure} 
\epsscale{1.2} 
\includegraphics[angle=0,scale=0.8]{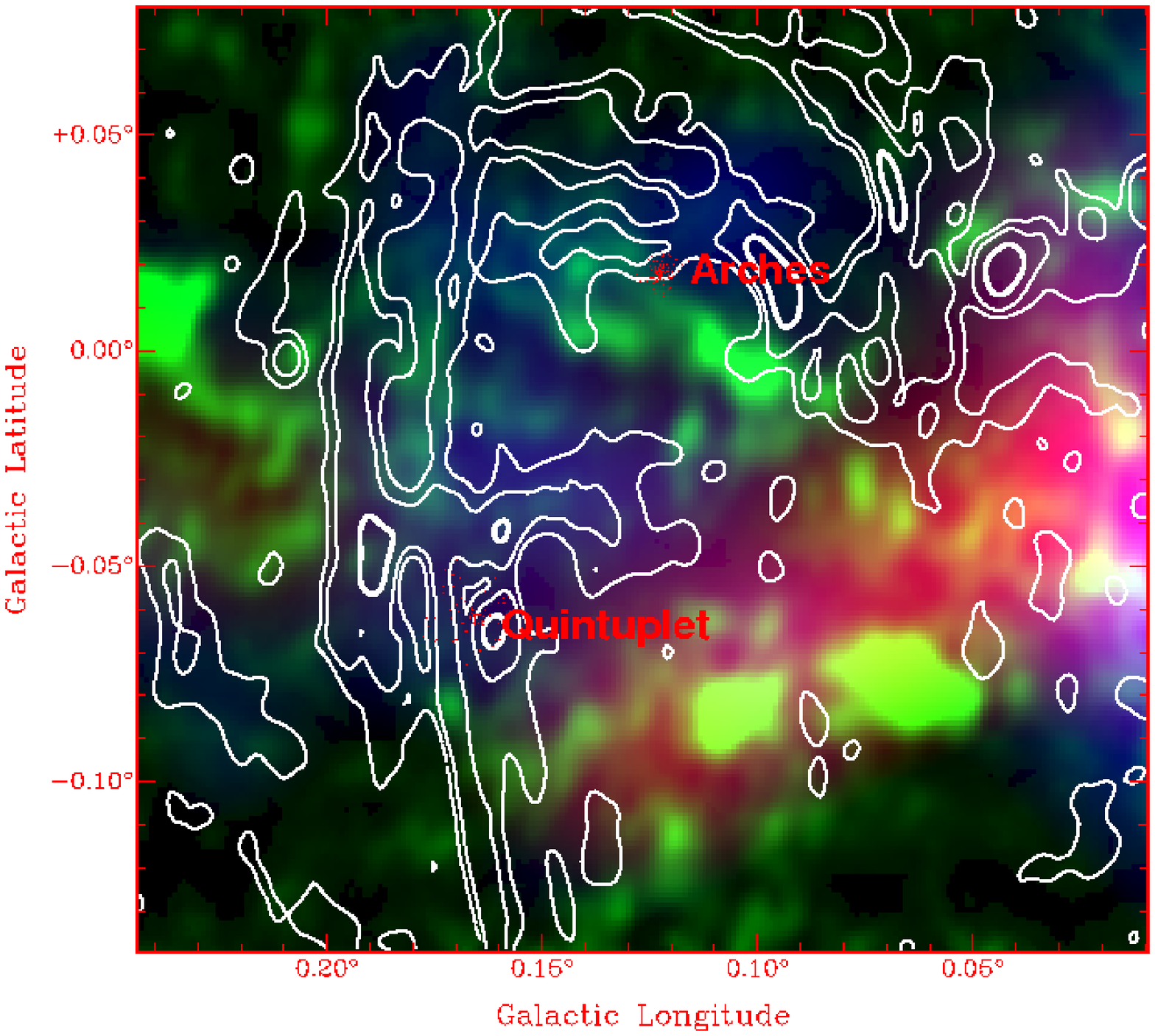} 

\caption{Multi--color composite of molecular gas close to the
  non--thermal filaments: {\it Red:} Integrated CO(4--3) emission,
  {\it Green:} Ammonia (1,1), {\it Blue:} \ion{C}{1}. Overlaid are the
  Arches and the Quintuplet stellar clusters in {\it red} contours and
  the 1.2\,cm radio continuum in {\it white} (see Fig.\,\ref{fig:m0}
  for the contour spacing). The CO and \ion{C}{1} data have been
  observed with the AST/RO south pole telescope
  \citep{mar04}. \label{fig:clusters}}
\end{figure}

\newpage
\begin{figure} 
\epsscale{1.2} 
\includegraphics[angle=0,scale=0.8]{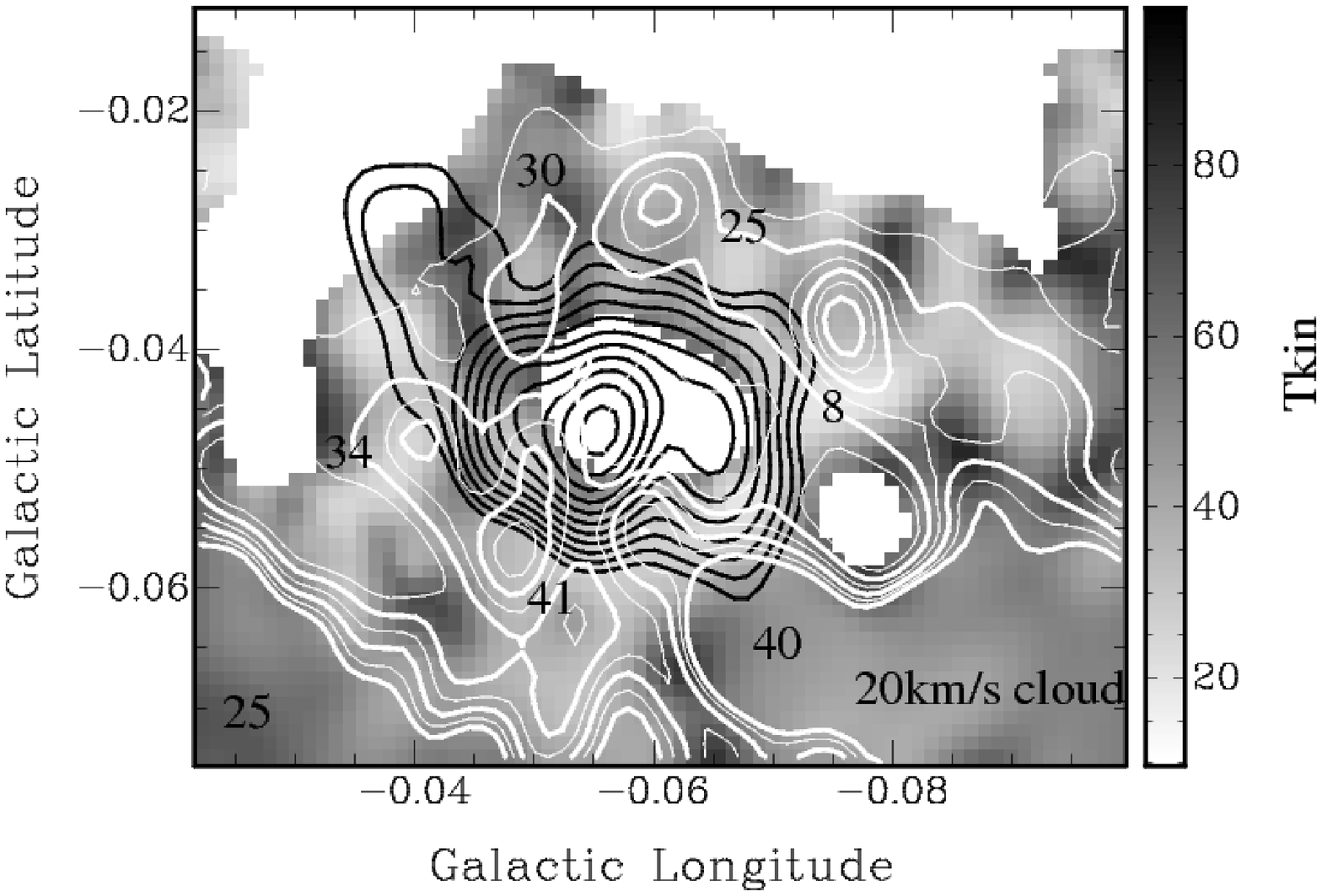} 

\caption{The Circumnuclear Disk around Sgr\,A*. Shown are the
  contours of the \amm(1,1) column density map in white contours
  (levels from $10\times10^{13}$\,\cden\ in $10\times10^{13}$\,\cden\
  increments), and the 1.2\,cm continuum outlining Sgr\,A* and the
  mini-spiral in black contours. The grayscale image is the kinetic
  temperature map. Numbers indicate the peak velocity value at
  various positions. The 20\,\kms\ cloud is located toward smaller
  Galactic latitudes. \label{fig:cnd}}
\end{figure}                                                                               

\end{document}